\documentclass[12pt,preprint]{aastex}
\eqsecnum
\accepted{}
\journalid{}{}
\articleid{}{}
\shortauthors{Boesgaard et al.}
\shorttitle{Abundances in M71 Turn-Off Stars} 
\received{2005 April 5}
\begin{document} 

\title{Chemical Composition in the Globular Cluster M71 from Keck/HIRES
Spectra of Turn-Off Stars}

\author{Ann Merchant Boesgaard\altaffilmark{1}}
\affil{Institute for Astronomy, University of Hawai`i at M\-anoa\\
 2680 Woodlawn Drive, Honolulu, HI {\ \ }96822-1839}
\email{boes@ifa.hawaii.edu}

\author{Jeremy R. King\altaffilmark{1}}
\affil{Department of Physics and Astronomy, 118 Kinard Laboratory,\\
Clemson University, Clemson, SC{\ \ }29634}
\email{jking2@ces.clemson.edu}

\author{Ann Marie Cody}
\affil{Department of Astronomy Option \\
 California Institute of Technology, MS 105-24, Pasadena, CA {\ \ }91125}
\email{amc@astro.caltech.edu}

\author{Alex Stephens\altaffilmark{1}}
\affil{Institute for Astronomy, University of Hawai`i at M\-anoa\\
 2680 Woodlawn Drive, Honolulu, HI {\ \ }96822-1839}
\email{alex\_c\_stephens@yahoo.com}

\author{Constantine P. Deliyannis\altaffilmark{1}}
\affil{Department of Astronomy, Indiana University\\
727 East 3rd Street, Swain Hall West 319, Bloomington, IN {\ \ }47405-7105}
\email{con@athena.astro.indiana.edu}

\altaffiltext{1}{Visiting Astronomer, W. M. Keck Observatory, jointly operated
by the California Institute of Technology and the University of California.}

%\clearpage

\begin{abstract} 

We have made observations with the Keck I telescope and HIRES at a resolution
of $\sim$45,000 of five nearly identical stars at the turn-off of the
metal-rich globular cluster M 71.  We derive stellar parameters and abundances
of several elements.  Our mean Fe abundance, [Fe/H]$=-0.80{\pm}0.02$, is in
excellent agreement with previous cluster determinations from both giants and
near-turnoff stars.  There is no clear evidence for any star-to-star abundance
differences or correlations in our sample.  Abundance ratios of the Fe-peak
elements (Cr, Ni) are similar to Fe.  The turn-off stars in M71 have
remarkably consistent enhancements of 0.2 - 0.3 dex in [Si/Fe], [Ca/Fe] and
[Ti/Fe] -- like the red giants.  Our [Mg/Fe] ratio is somewhat lower than that
suggested by other studies.  We compare our mean abundances for the five M 71
stars with field stars of similar metallicity [Fe/H] -- 8 with halo kinematics
and 17 with disk kinematics.  The abundances of the alpha-fusion products (Mg,
Si, Ca, Ti) agree with both samples, but seem a closer match to the disk
stars.  The Mg abundance in M71 is at the lower edge of the disk and halo
samples.  The neutron-capture elements, Y and Ba, are enhanced relative to
solar in the M71 turn-off stars.  Our ratio [Ba/Fe] is similar to that of the
halo field stars but a factor of two above that for the disk field stars.  The
important [Ba/Y] ratio is significantly lower than M71 giant values; the
pre-cluster material may have been exposed to a higher neutron flux than the
disk stars or self-enrichment has occured subsequent to cluster star
formation.  The Na content of the M71 turn-off stars is remarkably similar to
that in the disk field stars, but more than a factor of two higher than the
halo field star sample.  We find [Na/Fe] = +0.14 $\pm$0.04 with a spread less
than half of that found in the red giants in M71.  Excluding Mg, the lack of
intracluster $\alpha$-element variations (turn-off vis-a-vis giants) suggests
the polluting material needed to explain the abundance patterns in M71 did not
arise from explosive nucleosynthesis, but arose in a more traditional
$s$-process environment such as AGB stars.  The determination of light
$s$-peak abundances should reveal whether this pollution occurred before or
after cluster formation.

\end{abstract}


\section{INTRODUCTION}

It is important to determine the composition of unevolved stars in globular
clusters in order to learn the basic composition of the cluster from stars
that have not undergone any mixing with the interior layers.  We include as
``unevolved stars'' both main sequence and turn-off stars.  It is well known
that there can be large spreads in the abundance of some elements in the giant
stars of a given globular cluster (e.g. Kraft 1994, Sneden et al. 1997).  Most
previous high-resolution studies were limited to giant stars, whose
atmospheres and interiors are less well modeled than those of dwarf stars.
More extensive studies of the unevolved stars, which have little or no
advanced nuclear processing {\it in situ}, are expected to reveal the
clusters' original composition.

It now seems clear that mixing and evolutionary effects in cluster giants mean
that these stars cannot be used to probe the original abundance patterns in
clusters.  The the light elements (e.g. C, N, O, Na, Al) in the giants evince
rich intra-cluster differences that makes it impossible to ascertain a
universal pattern of light element alterations.  Good evidence exists, e.g.,
that CNO cycling and NeNa/MgAl burning in M13 has a considerable {\it in
situ\/} component (Pilachowski et al.~1996; Briley \& Cohen 2001).
Conventional wisdom now holds that such abundance modifications can be
explained by a combination of $p$-capture reactions (Denisenkov \& Denisenkova
1990; Langer, Hoffman \& Sneden 1993) and some non-standard deep mixing
mechanism (Cavallo, Sweigart \& Bell 1996) perhaps operating particularly more
efficiently in low metallicity clusters owing to smaller stellar interior mean
molecular weight gradients.  The most direct means, however, to explore {\it
in situ} versus primordial (or self-enrichment) mechanisms is to determine the
composition of unevolved stars in globular clusters.  More extensive studies
of the unevolved stars, which have little or no advanced nuclear processing
{\it in situ}, are expected to provide a baseline of clusters' red giants'
original composition(s) onto which the additions - by other mechanisms - are
made.

Several groups have begun studying globular cluster stars over a range of
evolutionary stages, including turn-off stars at high spectral-resolution
(e.g. Gratton et al.~2001; Cohen \& Mel\'endez 2005).  These initial efforts,
and previous photometric and low-resolution spectroscopic results, have
revealed strong evidence that some light element abundance alterations in some
clusters are present prior to reaching the red giant branch, thus suggesting
the possibilities of primordial variations, self-enrichment mechanisms, or
totally unexpected mixing mechanisms.  Briley et al.~(1994) found
anti-correlated CN and CH band strength variations at the turnoff of 47 Tuc,
and suggested that this cluster's CN distribution changes little with
evolutionary status; additionally striking is the existence of Na abundance
variations on the 47 Tuc main-sequence reported by Briley et al.~(1996).
Utilizing Keck data, Briley \& Cohen (2001) find that CNO abundances in M71
show little evolution from the main-sequence to bright giants; moreover, the
bimodal and anti-correlated CN and CH distributions of M71 red giants are
found to be in place on the main-sequence.  Gratton et al.~(2001) find that an
O-Na (and Mg-Al and C-N) anticorrelation, previously seen only in globular red
giants, exists in turn-off stars in NGC 6752.  Ram\'\i rez \& Cohen (2002)
(hereafter RC02) utilized high-resolution Keck/HIRES data to find that an O-Na
anticorrelation exists in M71 from its turnoff to giants above the red HB
level.  For an excellent review of globular cluster abundances and abundance
variations, see Gratton, Sneden \& Carretta (2004).

We have embarked on a high spectral-resolution study of turn-off stars in a
number of globular clusters.  Such stars are faint, so to observe them at high
resolution requires long integration times on the largest telescopes.  We have
used the Keck I 10-m telescope with HIRES to make observations of $V{\sim}18$
turn-off stars in M92, M71, M13, and M5.  Results on M92 have been published
on Li in Boesgaard et al.~(1998) and on Fe, Na, Mg, Ca, Ti, Cr and Ba in King
et al.~(1998).  The latter paper found [Fe/H]\footnote{We use the usual
notation [Element/H] = log N(Element/H)$_{star}$/log N(Element/H)$_{\odot}$} =
$-$2.52 for unevolved stars, compared to $-$2.25 for giants found by Sneden et
al.~(1991).  The reason for the dwarf-giant difference remains unclear, but
the homogeneous analyses of NGC 6397 and 6752 (Gratton et al.~2001) and M71
(Ram\'\i rez \& Cohen 2001, RC02) indicate no difference in [Fe/H] values (or
Fe peak, $n$-capture, and $\alpha$-element abundances for the M71 cluster)
between turnoff stars and more evolved cluster stars.  King et al.~(1998) also
found large Mg depletions and Na enhancements compared to HD 140283, a mildly
evolved halo subgiant with similar [Fe/H] and evolutionary state; [Ba/Fe] also
appeared to be enhanced in the M92 stars compared to HD 140283.  King et
al.~(1998) noted these qualitative patterns were reminiscent of those believed
to be due to MgAl and NeNa cycling in red giants.  (See the recent discussion
by Aikawa et al.~(2004) on the reaction rates in these chains related to
observations in red giants in globular clusters.)  Additionally, they
dedicated considerable discussion to wrestling with the difficulty in
understanding how such processing (whether {\it in situ} or from polluted
material processed in a previous stellar generation) could co-exist without
resulting in remarkably depleted Li abundances.  Bonifacio et al.~(2002) noted
a similar ``paradox'' for NGC 6397, which appears to be analogous to that seen
in the metal-poor field (Spite \& Spite 1986).  Bonifacio et al.~(2002)
mention preliminary evidence of prominent N enhancements in NGC 6397
subgiants, indicative of CN processing or supernovae in massive stars.  If
those enhancements are due to internal processing in their turnoff stars, then
how does Li remain ``normal''?  Similarly, Th\'evenin et al.~(2001) note that
one of their NGC 6397 turn-off stars demonstrates a relative 0.4 dex Na
deficiency; however, this star also evinces unremarkable Li.

This summary paints a complex and rich picture of globular cluster
abundances--one that will probably only be understood on a cluster-by-cluster
(if not star-by-star) basis.  Here we present observations, analysis, and
results from a study of five turn-off stars in M71.  We have determined
abundances for three iron-peak elements (Fe, Cr, and Ni), four alpha-fusion
elements (Mg, Si, Ca, and Ti), two rare-earth elements (Y and Ba) and Na.  The
globular cluster M71 is especially interesting for this work because it is
very old (${\ge}12$ Gyr; Grundahl, Stetson \& Andersen 2002), yet relatively
metal-rich ([Fe/H]${\sim}-0.8$) for a globular cluster.  Sneden et al.~(1994)
(hereafter SKLPS) find [Fe/H] = $-$0.79 from 10 red giants.  Cohen, Behr \&
Briley (2001), Ram\'\i rez et al.~(2001) and RC02 have reported on Keck/HIRES
observations of M71 of stars of a variety of evolutionary stages including
three stars near the turn-off.

The stars we observed in M71 have the same age, mass, luminosity, and
temperature as each other and as the stars in M 92, but the two clusters
differ in metallicity by a factor of $\sim$50.  We have selected M 71 because
1) it is a little brighter than M 92 at the turnoff ($V=17.7$ versus 18.0); 2)
despite considerable reddening ($E(B-V){\sim}0.28$), it has excellent
photometry (Hodder et al.~1992); 3) it is a disk globular with relatively high
metal content; 4) its space motion indicates that it is probably a member of
the thick disk population (Cudworth \& Hanson 1993).

We report on our four general goals: First, to characterize the abundances
of relatively unevolved stars in M71.  Second, to compare these with
determinations from both other near-turnoff stars (RC02) and red giants
(SKLPS, RC02) in M71.  Third, to compare our M71 turn-off abundances with those
in field halo stars and field disk stars of the same metallicity. Fourth, to
search for evidence of photospheric material having suffered $p$-capture and
deep mixing.

\section{Observations and Data Reduction}

Keck/HIRES spectral observations with respectable signal-to-noise (S/N) ratios
of stars near V = 18 require exposures of several hours.  However, individual
intergration times did not exceed 45-60 minutes due to the contamination by
``cosmic ray'' events.  On the other hand, the exposures needed to be long
enough to obtain a high enough signal so that individual frames could be
co-added reliably at the correct wavelength justification.  In metal-poor
stars where the spectral lines are weak this can be problematic.  Therefore,
we could only make these observations on cloudless nights with good seeing.

The spectra for this project were obtained at the Keck 10-m telescope with
HIRES (Vogt et al.~1994) over three observing seasons on four observing runs
comprising seven nights.  The same settings (slit dimensions, echelle and
cross-disperser angles, filters) were used on each run.  Approximately 2400
{\AA} were covered from 4430 -- 6880 {\AA} with some gaps between
orders longward of 5000 {\AA}.  The positions in the color-magnitude
diagram of our five stars can be seen in Figure 1.  The fiducial sequence is
taken from Hodder et al.~(1992).

The details of the observations for our five stars are shown in Table 1 which
gives the dates of the observations and the details of the multiple exposures
on multiple nights.  The star identifications are those of Hodder et
al.~(1992).  The photometric data for V and (B$-$V) are from Hodder et
al.~(1992) as is the assumed reddening correction of E(B$-$V) = 0.28.  This
value of the reddening agrees well with that of both Zinn (1980) of 0.27
$\pm$0.03 and that of Geffert \& Maintz (2000) of 0.27 $\pm$0.05.  The final
column shows the signal-to-noise ratio of the combined spectra.  As can be
seen in Table 1, the integration times were 135 - 285 minutes to reach the
listed S/N ratios (per pixel) of 50-60.  We have also made use of a spectrum
of the moon as a solar proxy that we obtained with the same settings on HIRES
as our stellar spectra.  The lunar spectrum was a 30 s exposure obtained on 29
July 1994 with a S/N ratio near 1400.

Standard data reduction procedures were carried out with IRAF\footnote{IRAF is
distributed by the National Optical Astronomical Observatories, which are
operated by AURA, Inc.~under contract to the NSF.} routines.  Individual
frames were trimmed, overscan-subtracted, and bias-subtracted (the median of
numerous overscan-subtracted bias frames from each night).  Typically 20
quartz flat-field exposures were obtained each night and these processed
frames were combined to form a master nightly flat-field frame.  Frames of the
stars in M71 were pre-processed in the same fashion and then divided by a
normalized flat field, which was produced by fitting a low order spline to the
blaze+lamp function.  

Scattered light was removed by fitting a low-order polynomial to the
inter-order regions across the dispersion and smoothing them along the
dispersion.  Cosmic ray hits were removed interactively.  Typically, 29 orders
were identified, traced and extracted.  The wavelength scale was found from
the positions of many lines in each order of the Th-Ar comparison spectrum.
The measured spectral resolution is $\sim$45,000.  The dispersion-corrected,
extracted 1-dimensional spectra were co-added order-by-order.  The final
spectrum of each star was then continuum normalized.

For most of the spectral analysis the individual orders were smoothed by a
3-pixel boxcar in order to increase the effective S/N ratio with minimal
impact upon the spectral resolution.  Examples of two spectral regions in
the spectra for three of the stars are shown in Figures 2 and 3.  Spectral
features from several ions are identified in these figures.  The three stars
shown all have the same value of (B$-$V) and the line strengths are clearly
similar in all three.

\section{ABUNDANCE ANALYSIS}

\subsection{Line Measurements}

We have made measurements of the equivalent widths of some 200 lines in each
of our five stars and 117 in the sun/moon spread throughout the 29 orders of
our Keck spectra.  The relevant features include those of the: Fe-peak
elements (\ion{Fe}{1} and \ion{Fe}{2}, \ion{Cr}{1} and \ion{Cr}{2}, and
\ion{Ni}{1}); alpha-fusion elements (\ion{Mg}{1}, \ion{Si}{1} and \ion{Si}{2},
\ion{Ca}{1}, and \ion{Ti}{1} and \ion{Ti}{2}); the alkali element Na
(\ion{Na}{1}); and the neutron-capture elements (\ion{Y}{2} and \ion{Ba}{2}).
The {\sf splot} package in IRAF was used to measure the equivalent widths via
gaussian profile fitting.  Lines that were noticeably blended were not used;
thus, nearly all could be measured using both sides of the line profile.  The
measured equivalent widths ranged from 7 to 200 m{\AA}, but lines used
for the parameter and abundance determinations were typically ${\le}80$ {\AA}.
Table 2 provides the species, wavelengths, lower excitation potentials,
oscillator strengths, and equivalent widths for all five stars and the sun.
The log gf values in Table 2 are from the compilations presented in Stephens
(1999) and Stephens \& Boesgaard (2002).

\subsection{Stellar Parameters and Uncertainties}

We have used the $(B-V)_0$ values from Hodder et al.~(1992) to determine the
stellar effective temperature of each star.  Although the reddening of 0.28 is
high, the good agreement with the values of Zinn (1980) and Geffert \& Maintz
(2000) suggest that it is rather well-determined.  Hodder et al. give errors
in B$-$V of 0.01 for stars in this V magnitude range.  We then use the
(B$-$V)$_0$ values to find T$_{\rm eff}$ for both the Carney (1983) (C83) and
the King (1993) (K93) temperature scales for metal-poor stars, as we have done
in earlier work (e.g.~Boesgaard et al.~1998).  In addition, we used
(B$-$V)$_0$ and a nominal metallicity of [Fe/H]$=-0.70$) to find T$_{\rm eff}$
values on the IRFM scale of Alonso et al.~1996); these temperatures agree
almost exactly with those on the C83 scale.  The uncertainty in the
temperatures comes from random errors in the photometry (0.01 mag), the large
reddening, and the calibrations.  Random errors of ${\pm}0.01$ mag in both the
photometry and reddening estimate both contribute an uncertainty of ${\pm}40$
K.  The two calibrations give $T_{\rm eff}$ values that systematically differ
by ${\sim}150$ K.  We carry through the detailed abundance analysis for both
sets of temperatures and their associated parameters: {\ }log g, [Fe/H], and
microturbulence ($\xi$).

We have used the newest Yale isochrones (Yi et al.~2001) to determine log g
values using an age of 12 Gyr, the median of the age values determined by
Grundahl, Stetson \& Andersen (2002), Salaris \& Weiss (1998), and Hodder et
al.~(1992), for both temperature scales for metallicities of both $-0.70$ and
$-1.30$ (and an interpolated value of $-1.00$).  Armed with a set of initial
temperatures, consistent theoretical gravities, and a range of overall
metallicities, we constructed model atmospheres characterized by these
parameters from the grids of Kurucz (1993).

Utilizing an updated (2000) version of the LTE analysis code {\sf MOOG}
(Sneden 1973), we computed absolute Fe abundances from both our \ion{Fe}{1} and
\ion{Fe}{2} equivalent width measurements for an array of trial
microturbulence values.  For these determinations we used only lines with
reduced equivalent width values log (W/$\lambda$)${\le}-4.82$ (generally
corresponding to equivalent widths ${\le}75-80$ m{\AA}).  For a given log g
and [Fe/H], we determined the combination of $T_{\rm eff}$ and $\xi$ that
resulted in the minimal slopes between Fe abundance and a) lower excitation
potential, and b) reduced equivalent width.  We also tested the effect of
varying log g on the agreement between Fe abundance as derived from
\ion{Fe}{1} and from \ion{Fe}{2}.

We adopted model parameters that are given in Table 3.  Although we have
carried out the abundance analysis throughout on both temperatures scales, we
have a preference for the ``cooler'' scale.  (This preference does not
imply a preference for the calibration used by C83 over K93 because
the large reddening for M71 makes the $B-V_0$ values are uncertain by about
$\pm$0.03.)  With the cooler scale: 1) there is better agreement with IRFM
temperatures, 2) the slopes for Fe I abundance vs excitation potential are
closer to zero with our spectroscopic temperatures, and 3) the match with the
new evolutionary tracks of Yi et al.~(2001) seems better as the tracks do not
reach the high turn-off temperatures of the hot stars on the hotter scale.

\subsection{Abundances}

We employed the updated LTE {\sf MOOG} code to determine abundances from all
our species.  MOOG's {\sf abfind} routine was used with the measured
equivalent widths in Table 2 and the appropriate Kurucz model atmosphere
characterized by parameters in Table 3.  The final abundances for each species
are given in Tables 4 (Fe), 5 (Na and $\alpha$ elements), and 6 (Fe-peak: Ni
and Cr, and n-capture: Y and Ba) and are discussed below in section 4.  For a
few lines of some species of particular interest we had to violate our
practice of only using features with log (W/$\lambda$)$<-4.824$ (e.g.
$\lambda$6141 \ion{Ba}{2}), but we stay within that limit for all the stellar
and solar lines of \ion{Fe}{1}, \ion{Fe}{2}, \ion{Ti}{1}, and \ion{Ti}{2}.
Where abundances were determined from lines from two ionization stages for an
element, an average was determined by weighting the results by the numbers of
lines measured.

Also included in those tables are the results for the sun from our lunar
spectrum from Keck/HIRES.  For these abundances we used the Kurucz solar
atmosphere and the lines for each species that appear in the last column of
Table 2.  Because the metallicity of the sun is considerably higher than that
of the M71 stars, we could not use exactly the same set of lines for the solar
abundance determinations.  This means that accurate $gf$ values might play a
role in determining accurate normalized abundances [X/H]--an issue which we
explore below.  However, there are a large number of well-determined solar
equivalent widths in the set we used for the stars.  The solar abundances also
appear in Tables 4, 5, and 6.  We have used the solar abundances that we
determined from the solar/lunar Keck spectrum for consistency in the
analysis.  Our results are in good agreement with the compilation of Grevesse
\& Sauval (1998), usually within 0.05 dex.  However, our Ca and Ni abundances
are 0.18 and 0.20 dex lower than the respective solar photospheric values of
Grevesse \& Sauval.

An abundance summary with abundances normalized to Fe is given in Table 7.  
The $\sigma$ values here represent the error in the mean.

\subsection{Uncertainties in the Abundances}

The internal uncertainty, ${\sigma}_{\mu}$, in the mean abundance for each
species in the five individual stars and the Sun are given in Tables 4, 5, and
6, along with the number of lines utilized in determining the mean abundance;
the former come from the line-to-line scatter for each species.  For those
species where only 1 or 2 lines were used, ${\sigma}_{\mu}$ could not be
determined.  In these cases, a conservative value had to be estimated for
later use.  In Table 8, internal uncertainties in the mean cluster abundance
ratios (column 3) are estimated from the star-to-star scatter of the
prenormalized abundances.  In column 4 we give the internal uncertainties in
the mean abundance ratios that additionally include contributions from the
solar values utilized in the abundance normalizations.  We note that, for
elements with multiple species, all these listed uncertainties include
contributions from each species weighted in the same fashion as the mean
abundances.

Table 8 also lists the sensitivities of the [Fe/H] and [X/Fe] ratios to the
changes in parameter values of T$_{\rm eff}$, log g, and $\xi$ given above
columns 5-7.  Typical errors are $\pm$50 K in T$_{\rm eff}$, $\pm$0.15 in log
g, and $\pm$0.15 in $\xi$.  Sensitivities include correlations of errors in
element X and Fe.  We note that all sensitivities take into account any
differences arising from utilizing different ions for a given element.
Sensitivity contributions from different species of a given element were
weighted in the same fashion as for the mean abundance determinations, i.e.~by
the number of lines used.  Based upon the photometric errors and our parameter
investigations, we believe that reasonably conservative estimates of random
1$\sigma$-level parameter uncertainties are ${\pm}50$ K, ${\pm}0.15$ dex, and
${\pm}0.15$ km/s in $T_{\rm eff}$, log $g$, and $\xi$, respectively.  This
means that total uncertainties in the mean abundance ratios are $0.05-0.10$
dex except for [Mg/Fe] and [Ba/Fe]; the latter two ratios have total
uncertainties of ${\pm}$0.12-0.13 dex.

We investigated the effects of our choice of $gf$ values by redetermining
abundances for each species utilizing only those specific lines for which
solar abundances abundances were also calculated.  The abundance differences
(``new'' minus ``old'') are given for each star in the fourth row of each
species in Table 4 for those species which utilized different line samples for
stellar and solar abundance determinations.  For example, had we used only
those \ion{Fe}{2} lines which were also measurable in the Sun, our
\ion{Fe}{2}-based mean stellar log N(Fe) value would have been 0.060 dex
lower.  Given the lower weight assigned to the \ion{Fe}{2} lines in
determining [Fe/H], however, this difference effects our final results
negligibly.  Indeed, none of our abundances is significantly affected by line
sample (and, thus, presumably $gf$ choice) except Na.  Table 5 indicates that
a truly consistent solar normalization of our \ion{Na}{1} lines would have
raised our final [Na/Fe] ratio by an additional ${\sim}0.08$ dex.  Whether
this systematic change yields a more accurate abundance compared to a value
derived from additional lines is, of course, unclear.  However, even the
difference for this one element is modest.

\section{Results}

\subsection{The M71 Iron Abundance}

The abundances for Fe for each star are given in Table 4.  For each ion we
give the 1$\sigma$ uncertainty and the number of lines used.  The
final abundance is the combined result obtained by weights equal to the number
of spectral lines that were used for the determination.  For Fe we have 79-86
Fe I lines and 8-11 Fe II lines for each star that give the final Fe/H.  There
are no significant star-to-star variations among the five stars so the mean
for the five - the cluster mean - is listed in the final column.  Also in
Table 4 are the results for the sun from our lunar spectrum with the
equivalent widths given in Table 2.  We used our solar values, obtained with
the same spectrograph, to find [Fe/H].

Our mean M 71 Fe abundance is [Fe/H]$=-0.80{\pm}0.06$ dex, where the
uncertainty includes internal uncertainties plus those in the parameters.  The
final column of Table 4 indicates that our \ion{Fe}{1}-based metallicity
([Fe/H]$=-0.81$) is in excellent accord with the \ion{Fe}{2}-based value
([Fe/H]$=-0.75$) within the internal uncertainties alone (0.03 and 0.07 dex
respectively). 

For these five turn-off stars we find $<$[Fe/H]$>$ = $-$0.80 $\pm$0.02, where
the uncertainty is just the error in the mean.  This is in excellent agreement
with [Fe/H] = $-$0.79 $\pm$0.01 from ten red giant stars in M71 of SKLPS.
These comparisons suggest that any intra-cluster dwarf-giant Fe discrepancy is
small.  This is in contrast to our previous study of the metal-poor globular
M92 (King et al.~1998), where we found Fe abundances to be nearly a factor of
two smaller in near-turnoff stars than deduced by others in M92 red giants.
One possible explanation may be the action of NLTE effects in the near turnoff
stars (King et al.~1998).  Th\'evenin \& Idiart (1999) have studied NLTE
effects on \ion{Fe}{1} lines in metal-poor stars.  Examination of their Figure
9 suggests that, at the LTE-based Fe abundance of M92 determined by King et
al.~(1998) ($-2.52$), NLTE corrections are indeed 0.3 dex.  Since these
corrections drop to ${\sim}0.1$ dex at our LTE-based M71 Fe abundance,
concordance with previous giant-based determinations is not remarkable.
Alternatively, as mentioned by King et al.~1998, it is possible that the M92
dwarf-giant difference is not real or is due to analysis differences.  Indeed,
Kraft \& Ivans (2003) suggest that their analysis of \ion{Fe}{1} and
\ion{Fe}{2} lines in M92 giants indicates the Th\'evenin \& Idiart (1999) NLTE
corrections are too large.  Kraft \& Ivans (2003) conclude that the
metallicity of M92 red giants is in the range $-2.50$ to $-2.38$, consistent
with the earlier estimate of $-2.52$ by Peterson, Kurucz \& Carney (1990) and
the King et al.~1998 turnoff results.

The agreement in [Fe/H] between the turn-off stars and the red giants in M71
is a confirmation that high quality spectra and model atmospheres can provide
the same benchmark abundance of Fe for this cluster.  We can now examine the
difference in other element ratios in giant vs.~turn-off stars to explore
whether nuclear processing and mixing have altered the abundances in the
giants.

\subsection{Overview of Abundances} 

The abundances of the other elements are given in Tables 5 and 6 and the mean
cluster abundances and parameter sensitivities are given in Tables 7 and 8.
Looking at the abundances of the various species in our five M71 stars, there
is no firm evidence for genuine star-to-star differences; in particular, no
star stands out has having consistently deviant abundances given the
uncertainties.  Depending on the species, we infer that any genuine
star-to-star differences must be at a level of ${\le}0.05-0.1$ dex for the
particular elements in our sample of stars.

\subsubsection{Comparison with Halo Dwarfs of Similar Fe Abundance}

Our first comparison is between our M71 stars and field stars of similar
metallicity that have been analyzed by Stephens \& Boesgaard (2002)
(hereafter SB02) because we have used the same line list and the same
procedure.  There are eight stars in that sample with [Fe/H] between $-$0.6
and $-$1.2.  The abundances in these eight stars are shown in Table 9.  Figure
4 shows the abundances of these Fe-normalized abundances with our mean for M71
as a function of [Fe/H] for four alpha-elements, two Fe-peak elements and two
n-capture elements.  Compared to the field halo stars, M71 is high in Ca (but
would be in the midst of the halo star points at +0.14 had we used the
Grevesse \& Sauval (1998) solar Ca of 6.36).  Although the field stars appear
to be low in Ni, which is discussed by SB02, the M71 point would fall among
them at $-$0.10 with the Grevesse \& Sauval (1998) solar Ni of 6.25.  The
field stars are also apparently low in Na (see below).

\subsubsection{Comparison with Edvardsson et al.~low metallicity disk stars}

Edvardsson et al.~(1993) (hereafter EAGLNT) derived abundances of 13 elements
in some 189 field F and G dwarf stars in the Galactic disk.  Several of these
stars have [Fe/H] values similar to that of M71.  We have separated out those
stars with [Fe/H] within $\pm$0.20 dex of $-$0.80 to compare with our M71
turn-off mean abundances; there are 17 such disk stars and their abundances
are given in Table 9.  Figure 5 is a similar plot to Figure 4 with the disk
stars as filled circles; EAGLNT did not determine [Cr/Fe] so the lower left
panel in Figure 5 is [Na/Fe] rather than [Cr/Fe] as in Figure 4.

We can compare Figures 4 and 5 of the abundances in the halo field stars and
in the disk dwarfs with our M71 results.  Note that although the samples of
the disk stars of EAGLNT and those of the halo stars of SB02 have similar
metallicities, the orbital characteristics are very different in the two
samples.  We have examined the U, V, W velocities, the perigalacticon
distances, R$_p$, and the maximum distance above the Galactic plane,
Z$_{max}$, for all the stars in each sample.  For example, the $|$U$|$ values
are 4 - 123 km s$^{-1}$ in the disk stars, but 180 - 358 km s$^{-1}$ in the
halo stars.  With the exception of one disk star, Z$_{max}$ is $<$ 0.85 kpc
for the disk stars, while the halo star sample has Z$_{max}$ values are $>$1.2
kpc, except for one star with low Z$_{max}$ but with $|$U$|$ = 254 km
s$^{-1}$.

The M71 means for the $\alpha$-elements appear to match the abundances of the
disk dwarfs of comparable metallicity better than they match those of the halo
field stars.  The value of [Mg/Fe] is at the lower part of the range for both
the disk stars and the halo stars, while [Si/Fe], [Ca/Fe], and [Ti/Fe] are
somewhat higher in M71 than in the halo stars.  According to Cudworth (1993)
the metallicity and U, V, W velocities of M71, $-$76, $-$60 and $-$3 km
s$^{-1}$, are characteristic of the thick disk population.

The abundance of the n-capture element, Ba, in M71 is well-matched by the
abundances in the halo stars, but the mean value is more than a factor of two
higher than [Ba/Fe] in the disk stars.

\subsubsection{Comparison with Sneden et al.~M71 Giants}

Figure 6 shows the elemental abundances normalized to Fe as a function of
atomic number and includes values for our M71 turn-off stars, the ten M71 red
giants of SKLPS, and the halo field stars.  The giants and turn-off stars
agree well in Na and Ni.  The alpha-element Ca seems high in the turn-off
stars, but would fit perfectly at +0.14 with the Grevesse \& Sauval (1998)
solar Ca.  The alpha-elements Ti and Si are somewhat lower in the turn-off
stars than in the giants.  Here our Ni abundance is in excellent agreement
with the values from the SKLPS giants.

\subsubsection{Comparison with Ram\'\i rez \& Cohen}

There are three RC02 M71 stars that are similar to ours in
temperature ($\sim$5800 K) and log g ($\sim$4.05) values, which they class as
turn-off stars.  Although the S/N (per pixel) at 46, 44, 32 for their three
stars are somewhat lower than ours for our five stars, their abundances
provide a reasonable comparison for our own results.  If we determine the mean
[Fe/H] from those three stars the same way as we have done for ours,
i.e. weighing the results from Fe I and Fe II by the number of lines used, we
determine $<$[Fe/H]$>$ = $-$0.76 $\pm$0.05, which is the same within the
errors as our mean of $-$0.80 $\pm$0.02.

Figure 7 shows the mean abundances normalized to Fe for our five turn-off
stars and the three from RC02 along with the abundances of 19 red giants from
RC02.  One sigma error bars are shown which are the error in the mean.
Generally, the abundances agree to within 1-2 $\sigma$ for the turn-off stars.
An exception is [Si/Fe], but although their two stars agree with each other,
each has errors of $\pm$0.13 dex.  They were only able to measure Cr from one
line in one star, so that discrepancy is not important; our Cr abundances come
from 16-19 lines of Cr I and Cr II in all five stars.  Their Ni abundance is
from 2-7 lines in three stars, while ours is from 15-18 lines in five stars.
Similarly, we have 21-25 lines of Ti I and Ti II for our Ti abundance in the
five stars versus their 4-6 Ti I lines in three stars.  Inasmuch as we
concentrated on five turn-off stars, we could measure 2-5 times as many lines
as they did for most species, i.e. our turn-off abundances are are based on
much more data.  However, they examined a large range in stellar luminosity
and evolutionary stage and determined abundances of 23 elements, many more
than our sample of 10 elements.

It is interesting to compare their results for 19 red giants with those of
SKLPS for ten red giants and for their red giants with our turn-off stars.
The agreement for the two sets of red giants is good for Na, Si, and Ni.  The
situation for Ca and Ti seems confusing (see discussion below about
alpha-elements).  Their red giants seem to be enhanced in Mg and, to a lesser
extent, in Na compared to our turn-off-star abundances.

\subsection{Fe-Peak Elements}

The other two Fe-peak elements, Cr and Ni, are very similar to Fe in our
M71 stars.  We derive abundances of [Cr/H] of $-$0.75 $\pm$0.04 from 16-19 Cr
I and Cr II lines per star and [Ni/H] $-$0.71 $\pm$0.04 from 15-18 Ni I lines
per star.

The Fe normalized means for the cluster are [Cr/Fe] = +0.05 $\pm$0.04 and
[Ni/Fe] = +0.09 $\pm$0.04.  We can compare our results for Ni with those of
SKLPS which are based on 3-5 Ni lines in ten giants in M71 (see Figure 6).
(They did not measure Cr.)  Their value of $<$[Ni/Fe]$>$ = 0.07 $\pm$0.04
(with $\sigma$=0.11) is the same as ours to within the errors.  The 19 red
giants of RC02 have $<$[Ni/Fe]$>$ = +0.02 $\pm$0.01 (with $\sigma$=0.04) from
13-38 Ni I lines.  The similarity of the abundances derived for our turn-off
stars and the two sets of giant stars for Ni indicates a constancy between the
evolved and unevolved stars.  In the case of Cr RC02 find a lower
$<$[Cr/Fe]$>$ = $-$0.13 $\pm$0.01 (with $\sigma$=0.06) from 3-10 Cr I lines
compared to our +0.05 $\pm$0.04 (see Figure 7).  (Their [Cr/Fe] for their one
turn-off star is still lower at $-$0.17.)  Our [Cr/Fe] (+0.05 $\pm$0.04) is
the same to within the errors of the halo stars of SB02 (+0.02 $\pm$0.02).

\subsection{Sodium}

We have determined Na abundances in each star from 3-4 Na I lines.  These
results appear in Table 5.  There appear to be no variations from one star to
the next.  Our mean ratio [Na/Fe] = +0.14 $\pm$0.04 agrees well with the
$+0.20{\pm}0.05$ value from RC02 from 3 Na I lines for three turn-off stars
just considering our internal uncertainty alone.

Figure 8 shows our M71 results for [Na/Fe] compared to those of the halo field
stars of SB02 ($<$[Na/Fe]$>$ = $-$0.23 $\pm$0.13) and the disk dwarf stars of
EAGLNT ($<$[Na/Fe]$>$ = 0.10 $\pm$0.01).  The Na content of the M71 turn-off
stars is remarkably similar to that of the disk dwarfs and more than a factor
of two higher than that in the halo stars with similar [Fe/H].

Our [Na/Fe] of +0.14 for the turn-off stars has a spread of +0.01 to +0.23,
less than half that found in the red giants by both SKLPS and RC02.  The
spread in our turn-off stars is consistent with the uncertainties in the
determinations of [Na/H] for the individual stars and provides no indication
of an intrinsic spread.  In their study of 10 red giants in M71, SKLPS derived
Na abundances from 4 Na I lines which showed a spread in [Na/Fe] from -0.04 to
+0.53 and a mean of +0.24 $\pm$0.06 (with $\sigma$=0.18).  Similarly, RC02
find $<$[Na/Fe]$>$ = +0.24 $\pm$0.03 (with $\sigma$=0.14) and a range of +0.04
to +0.57 in their 19 red giant stars.  As can be seen in Figures 6 and 7 there
is little difference in the {\it mean} [Na/Fe] content between the M71
turn-off stars and the M71 red giant stars.  However, the sizeable spread in
the abundances for individual giant stars indicates that some of the giant
stars have Na enhancements, as found in several globular clusters (e.g.~in M3
by Kraft et al.~1992, in M13 by Kraft et al.~1993, 1997).

An increased Na could result from the capture of a proton by $^{22}$Ne to form
$^{23}$Na; this could occur in the H-burning shell.  The resultant Na is
dredged up to the photosphere where it can be observed.  An {\it in situ}
processing scenario may or may not be responsible for the cluster's mild O-Na
anti-correlation or the range in Na enhancements in the red giants.

Figures 9a and 9b show comparisons of Na with Mg for M71 and the two sets of
field stars.  Whereas Mg seems similar to the halo stars, the ratio of [Na/Mg]
of +0.05 is higher than that in any of the halo or disk field stars. (One can
see the M71 point as an extension of the disk-star points toward higher
[Na/Mg] at lower [Mg/H]; this is analogous to the trend of [O/Fe] vs.~[Fe/H],
but with Mg as the chronometer rather than Fe.)  For the mean of the 19 red
giants, RC02 find [Na/Mg] = $-$0.12 from an apparent enhancement in Mg in the
giants.

\subsection{Alpha Elements}

We have calculated abundances for four alpha-elements: Mg, Si, Ca, and Ti.
For Mg there are only 2-3 lines that we can use for the abundance
determination, only 4 for Si, but 18-20 for Ca and 21-25 for Ti.  For both Ti
and Si we have used lines from neutral and ionized states.  The abundances for
these $\alpha$-fusion elements, along with the 1$\sigma$ uncertainty and the
number of lines used in the abundance calculation are presented in Table 5.
The Fe-normalized ratios are in Table 7 along with the weighted mean of
[$\alpha$/Fe].  (The weights are the number of lines used in the determination
of the abundance of each element.)  Again we see no significant star-to-star
variations so we have found a cluster average for turn-off stars for each
element, both [$\alpha$/H] and [$\alpha$/Fe].  The average alpha-enhancement
is [$<\alpha>$/Fe] = +0.29 $\pm$0.05.

Our $\alpha$-element ratios [Si,Ca,Ti/Fe] show a clear overabundance given the
mean values of +0.22, +0.32, and +0.30 with typical total internal
uncertainties of 0.05 dex.  RC02 find these same ratios from their 2-3
turn-off stars of +0.01, +0.40, and +0.16.  For their 19 red giants they find
+0.25, +0.43, and +0.21 with uncertainties of typically 0.02 dex.  The red
giants of SKLPS have those ratios as +0.31, +0.14, and +0.48 with
typical uncertainties of 0.04 dex.  The straight means for these three
$\alpha$ elements are +0.30 (us), +0.19 (RC02 T.O.), +0.29 (RC02 giants), and
+0.31 (SKLPS); this is a remarkably consistent alpha enhancement in this
cluster in both turn-off and giant stars.  

Our mean [Mg/Fe] is +0.09.  This is lower than the RC02 value of +0.28 (based
on only two turn-off stars they measured) and lower than the RC02 value of
+0.37 (based on 19 red giants).  It is also lower than the results of Shetrone
(1996) who found +0.32 $\pm$0.04 from eight M71 giants.  However, our internal
uncertainties for Mg are larger than for our other elements.

It is possible that our [Mg/Fe] is in accord with the delayed rise in [Mg/Fe]
suggested by Fuhrmann et al.~(1995; see their Figure 5) and implies a genuine
dwarf-giant difference.  Further, the NLTE reanalysis of Galactic Mg ratios by
Idiart \& Th\'evenin (2000) suggests there is significant scatter ($\sim$0.6
dex) in [Mg/Fe] at [Fe/H] $\sim-$0.8.  While we have not done the NLTE
calculations here, we surmise that our M71 turn-off stars with [Mg/Fe] $\sim$
+0.1 simply lie in the midst of this scatter.

\subsection{N-Capture Elements}

Table 6 contains the abundance results for our five M71 turn-off stars for Y
and Ba based on 3 lines of Y II and 2 of Ba II and Table 7 gives the mean for
each element normalized to Fe.  We find and [Ba/Fe] = +0.23 $\pm$0.06, [Y/Fe]
= +0.15 $\pm$0.05 with [Ba/Y] = +0.08.

Figures 4 and 5 show how our results compare to field stars of similar
metallicity in the halo (SB02) and in the disk (EAGLNT).  Our M71 stars are
quite similar to the halo stars in [Ba/Fe] where the halo mean is +0.16
$\pm$0.01, but a factor of 2 larger than the disk stars where $<$[Ba/Fe]$>$ =
$-$0.08 $\pm$0.01.  For [Y/Fe] our mean of +0.15 is higher than the value of
$-$0.03 found for both the halo stars and the disk stars.  These enhancement
are modest, and only marginally significant given the uncertainties, but may
simply be the consequence of exposure of pre-cluster material to slightly
higher neutron flux than that received by most disk field stars.  The halo
stars have [Ba/Y] = +0.19, while the disk stars [Ba/Y] = $-$0.05 compared to
our intermediate value of +0.08.  These heavy-to-light ratios, a measure of
neutron exposure, are certainly not in conflict with this idea.

The three M71 turn-off stars studied by RC02 have no Y determinations and show
a large range in [Ba/Fe]: +0.22, +0.33 and +0.61.  The quoted error for the
third star is $\pm$0.32; the mean for the other two is +0.27, similar to our
[Ba/Fe] of +0.23 $\pm$0.06.  Their three evolved stars which do have
determinations of [Y/Fe] yield $-$0.22 while the 25 cluster stars of all
evolutionary phases give [Ba/Fe] = +0.34.  Their [Ba/Y] of +0.56 is a high
value, comparable to those seen in some Ba, S, CH, stars and ${\omega}$ Cen
giants.  Such large ratios can be produced by very large neutron exposure,
which yields significant production of heavy-$s$ elements (e.g., Ba) at the
expense of light-$s$ elements (e.g., Y and Zr) as indicated by both early
Clayton et al.~(1961) and more recent theoretical calculations (figure 5 of
Busso et al.~1995).  The expectation from the $s$-process calculations is that
the adjacent light-$s$ element Zr should show an abundance comparable or just
slightly larger than Y, and this is seen in the results of RC02 whose
cluster results indicate [Y/Zr]${\sim}-0.09$.  This ratio is unremarkable
compared to field stars of similar metallicity (Figure 29 of Edvardsson et
al.~1993; Figure 5 of Gratton \& Sneden 1994); thus, the RC02 Zr
results seem to confirm their Y results.

These heavy elements have been studied in globular clusters and field halo
stars by James et al.~(2004).  For their five field stars with [Fe/H] similar
to $-$0.80 for M71, we find [Ba/Fe] = +0.29, [Y/Fe] = $-$0.05, and [Ba/Y] =
+0.34.  The difference is mainly in the low [Y/Fe] values.  For the three
turn-off stars in the globular cluster 47 Tuc, which has [Fe/H] = $-$0.69,
they find [Ba/Fe] = +0.22, [Y/Fe] = +0.06, and [Ba/Y] = +0.16, values very
similar to ours.

We can conclude that the primordial material from which M71 formed underwent
at least modestly larger neutron exposure than local field disk stars of
similar [Fe/H].  This appears to be true in 47 Tuc as well as deduced from
their heavy element abundances in turn-off stars.

The 0.6 dex [Ba/Y] difference between our M71 turn-off stars and the cluster
giants of RC02 is surprising, and has potentially significant implications for
cluster self-enrichment as manifested by the effects of convective dilution.
The reality of this difference needs confirmation by additional observations
of light s-peak elements (Y, Rb,Sr, Zr) in M71.  In this regard, we note the
intriguing circumstance that the mean [Zr/Fe] ratio (-0.23 +/- 0.09) of RC02's
7 most highly evolved stars (log g <=1.75) is lower than that (+0.10+/-0.13)
of their 3 least-evolved stars (log g>=2.70) with Zr measurements at the
2.1 sigma level.

\subsection{M71 and ${\alpha}$ Anomalies in the Bulge and Disk}

In their study of 10 M71 {\it giants}, SKLPS noted ``normal'' 0.2-0.3 dex
enhancements of [Si/Fe] and [Ca/Fe].  These match our near-turnoff ratios
well.  However, they note that their Ti ratio is unusually large:
[Ti/Fe]${\sim}+0.5$, and call attention to the fact that a similar dichotomy
between [Ti,Mg/Fe] and [Si,Ca/Fe] is seen in moderately metal-poor disk and
bulge giants (e.g., McWilliam \& Rich 1993) {\it and\/} in some disk dwarfs
EAGLNT.  However, both our [Ti/Fe] ratios and those of RC02 do not confirm the
larger Ti overabundances noted by SKLPS; rather, they suggest unremarkable
values of $+0.2$ to $+0.3$ indistinguishable from Si and Ca.

\subsection{Self-Enrichment and Pollution Redux} 

In our previous Keck/HIRES work on M92 (King et al.~1998), we suggested that
pollution of these stars' atmospheres along with later {\it in situ\/}
processing was needed to understand the global picture of this cluster's
abundance patterns.  Whether such pollution was in the form of abundance
variations or previously processed material in the proto-cluster gas or
whether it happened after (at least the first generation of?)  cluster stars
formed (self-enrichment) was not clear.  While the idea of such pollution and
intra-cluster abundance variations from a {\it non in situ} source may not be
easy to accept, we note additional support for this picture in M92.  In their
NLTE reanalysis of numerous literature Mg and Ca data, Idiart \& Th\'evenin
(2000) include our M92 star equivalent widths.  Their Figure 3 contains the
striking result that the NLTE [Ca/Fe] abundances in six M92 stars fall nearly
precisely along 3 bifurcated sequences defined by field star data.  While
significant scatter is present for Mg field star data, they suggest a similar
trichotomy for the NLTE [Mg/Fe] ratios.

As suggested by Idiart \& Th\'evenin (2000) these intra-cluster differences
may indeed be similar to those now well-known in ${\omega}$ Cen, and presumed
to arise from self-enrichment from Type II supernovae.  Moreover, the
correspondence of the M92 star [Ca,Mg/Fe] ratios and the field star sequences
may provide renewed support for the idea of globular clusters as the original
reservoirs of halo field stars.  Regardless, investigating the correspondence
of M 71 [Ca,Mg/Fe] ratios and the sequences defined by field stars is clearly
important.

However, it is not yet clear that this important task is possible.  At
[Fe/H]${\sim}-0.8$, the field star [Mg,Ca/Fe] sequences in Figure 3 of Idiart
\& Th\'evenin (2000) are separated by a modest 0.05-0.1 dex, which is
comparable to or smaller than the internal uncertainty in our star-to-star Mg
abundances.  Moreover, our data need to be subjected to a NLTE analysis in an
uniform as possible manner compared to Idiart \& Th\'evenin (2000).  At
present, we can only note that there is no strong evidence of intra-cluster
[Ca/Fe] differences in M71 from our data; the largest difference seen between
our five stars is 0.16 dex, which is a 3${\sigma}$ difference based on the
internal star-to-star uncertainties alone; the [Ca/Fe] ratios for the three
turn-off stars of RC02 show no evidence of statistically significant
bifurcation either.  Magnesium abundances of additional little-evolved M71
stars using even higher-quality data would be of great interest in pursuing
unsettled questions regarding star-to-star differences perhaps indicative of
cluster self-enrichment or primordial abundance variations.

While it was unclear from the work of King et al.~(1998) whether any {\it non
in situ\/ } abundance variations in M92 were primordial or the product of
cluster self-enrichment, they noted the potential importance of turnoff-giant
comparisons in addressing this issue.  If self-enrichment occurs after stars
have formed, any products that are enhanced (at levels which are easy to
detect) in the near turnoff stars may not appear to be enhanced in the red
giants.  The turn-off stars have relatively thin surface convection zones and
the enhancements there may subsequently become unrecognizably diluted as these
stars (or stars just slightly more massive) develop deep surface convection
zones as they ascend the red giant branch.  King et al.~(1998) noted that the
${\sim}0.5$ dex lower abundance for [Mg/Fe] in their M92 near turnoff stars
compared to the M92 giants in Shetrone (1996) might indicate the action of
self-enrichment in the giants.  The difference between the mean M92 near
turnoff NLTE [Mg/Fe] ratios from Idiart \& Th\'evenin (2000) and Shetrone's
(1996) giant mean remains large at $0.43$ dex.

Comparison of intra-cluster dwarf-giant abundances of $\alpha$-elements is an
efficient manner to search for self-enrichment signatures since the production
of these is canonically presumed to occur in explosive supernovae having
massive (i.e., short-lived) progenitors.  We note that, in principle, such
stars could also be responsible for $r$-process contributions to a possible
dwarf-giant $n$-capture difference noted above; while we refer to our heavy
elements as ``$s$-process'' ones, it must be remembered that such designations
are based on {\it solar system\/} data.  In M71, the RC02 results for Ti and
Ca do not provide any evidence for significant near turnoff-giant differences
that might be taken as evidence of self-enrichment.  Our near-turnoff [Mg/Fe]
ratio of $+0.09{\pm}0.03$ is considerably smaller than the value of the RC02
evolved stars ($+0.39{\pm}0.014$), which itself should be raised (or ours
lowered) by 0.04 dex to account for the modest Fe difference between our
studies.  While interesting, the meaning of this difference is unclear since
the two turn-off stars from RC02 give [Mg/Fe]$=+0.28{\pm}0.09$, in accord with
their giant results.  Our turn-off [Si/Fe] ratio is +0.22 $\pm$0.04,
essentially the same as that of RC02 for evolved stars of $0.25{\pm}0.03$ dex.

We conclude that no $\alpha$ abundance difference is clear between the giants
and the turn-off stars.  If the $n$-capture difference noted above is real and
resulted from self-enrichment, then we surmise that polluting material arose
in a traditional $s$-process environment--though perhaps one providing an
unusually large neutron fluence.  We reiterate that mapping any variation of
light $s$-element variation with evolutionary status in M71 is critical to
determining whether such pollution happened before cluster formation or via
self-enrichment after star formation in the cluster.

\section{Summary and Conclusions}

We have derived abundances of numerous elements in five stars near the turnoff
of the relatively metal-rich globular cluster M71 via high-resolution
($R{\sim}45,000$) spectroscopy with S/N of 50-60 per pixel acquired with
Keck/HIRES.  Our five stars are very similar to each other in temperature,
luminosity, mass, and age.  Our derived Fe abundance, [Fe/H]$=-0.80{\pm}0.06$
is in excellent accord with the values measured in both giants and
near-turnoff stars by Sneden (1994) and RC02.  The lack of a dwarf-giant Fe
discrepancy like that we found in the metal-poor cluster M92 may reflect the
reduced effects of NLTE on the \ion{Fe}{1} abundances in the metal-rich case
as suggested by the work of Th\'evenin \& Idiart (1999).

Within our own limited data set, we find no evidence of star-to-star abundance
deviations in any element.  Furthermore, there are no stars that are
consistently higher or lower than the others in the abundances of the various
species we studied; among other things, this indicates that our relative
temperatures are correct.

We have compared our abundances with two field star samples with metallicities
similar to that in M71: eight halo stars from SB02 and 17 disk stars from
EAGLNT.  Our $\alpha$-product abundances, Mg, Si, Ca, Ti, seem to fit with
both the disk and the halo stars, but perhaps a somewhat better match with the
disk stars.  (The space motion of M71 is similar to that of disk stars
according to Cudworth \& Hanson (1993).)  For Mg the M71 mean is near the
lower edge of both samples.  Our M71 mean is more similar in Ni to the disk
stars, but Cr matches the halo sample well.  Whereas the light s-process
element Y is enhanced in M71 relative to both field stars samples, the heavier
s-process element Ba is like the halo stars and enhanced more than a factor of
2 above the disk stars.

All the $\alpha$ elements we observed (Mg, Si, Ca, TI) are enhanced relative
to Fe as compared to the Sun.  The mean enhancement of $<$[$\alpha$/Fe]$>$ is
+0.29 $\pm$0.05.  This is strikingly similar to the mean enhancements in those
elements in the red giants in M71 studied by SKLPS and RC02.

It is known that M71 evinces O-Na anticorrelations down to the main-sequence
turn-off (RC02) that are presumed to arise from $p$-capture reactions.  The
totality of data now available for M71 suggests to us the following two
conclusions.  First, the similarity of our turn-off Na abundances, as well as
those of RC02, with the values observed in M71 giants (SKLPS; RC02) indicates
that this $p$-capture mechanism is not an {\it in situ} process unless there
is a glaring fundamental failure in our basic understanding of stellar
structure.  The atmospheres of M71 stars must be polluted by material having
undergone such processing elsewhere.  Second, the lack of clear evidence for
intracluster $\alpha$-element variations suggests that this processing was not
the result of self-enrichment via explosive nucleosynthesis -- a process which
could, in principle, also affect the heavy elements if the $r$-process came
into play.  The M71 [Ba/Fe] ratio suggests that material in M71 stars once was
present in an $s$-process environment of at least modestly higher neutron
fluence compared to nearby field stars of similar [Fe/H].  It remains unclear
whether the $s$-process polluting source acted prior to cluster formation or
after (via self-enrichment).  If there is a genuine dwarf-giant Y and/or Zr
difference in M71, then it would suggest that a source of modest neutron
fluence self-polluted M71 (to produce high light $s$ abundances perhaps seen
near the turnoff) in addition to a very high neutron fluence source acting
prior to cluster formation (to produce the high heavy-to-light $s$ ratios
perhaps seen in the giants).  Consistent determination of light $s$-elements
(Rb, Sr, Y, Zr) in a considerably larger sample of M 71 turn-off, subgiant,
and giant stars is needed to address this critical issue; the possibility of
intra-cluster Mg variations also merits additional study.

Finally, we note the recent work on elemental abundances in the globular
clusters M4 ([Fe/H]$=-1.08$) and M5 ([Fe/H]$=-1.21$) by Ivans et al.~(1999,
2001).  Ivans et al.~(2001) call attention to the differences in these
clusters based upon the morphology of the O-Na anticorrelation patterns and HB
morphology.  In particular, they note the similarity in the light element
processing patterns between M4 and M71, as well as the HB morphology.  Indeed,
it might be hoped that such an association is a meaningful classification--
perhaps related to the dominance of {\it in situ} processing versus pollution,
or perhaps simply related to metallicity since this parameter affects the
molecular weight gradient relevant for deep mixing in stellar interiors as
well as serving as the ``first parameter'' controlling HB morphology.  The
factor of ${\ge}2$ difference in the M4 and M71 [Ba/Fe] ratios, however,
reminds us that the detailed chemical evolution histories of globular clusters
are individually unique.  As is clear from this investigation, unraveling
these histories will require continued study on a star-by-star and
cluster-by-cluster basis involving determination of abundances in near turnoff
stars via high-resolution spectroscopy on current and future generations of
large aperture telescopes.

\acknowledgments

This work has been supported by an NSF award to AMB, AST-0097955, and the NSF
REU program award to the Institute for Astronomy at the University of Hawaii,
AST-9987896.  JRK gratefully acknowledges support for this work from NSF
awards AST-0086576 and AST-0239518 as well as a generous grant from the
Charles F.~Curry Foundation to Clemson University.

\clearpage

\begin{figure}
\plotone{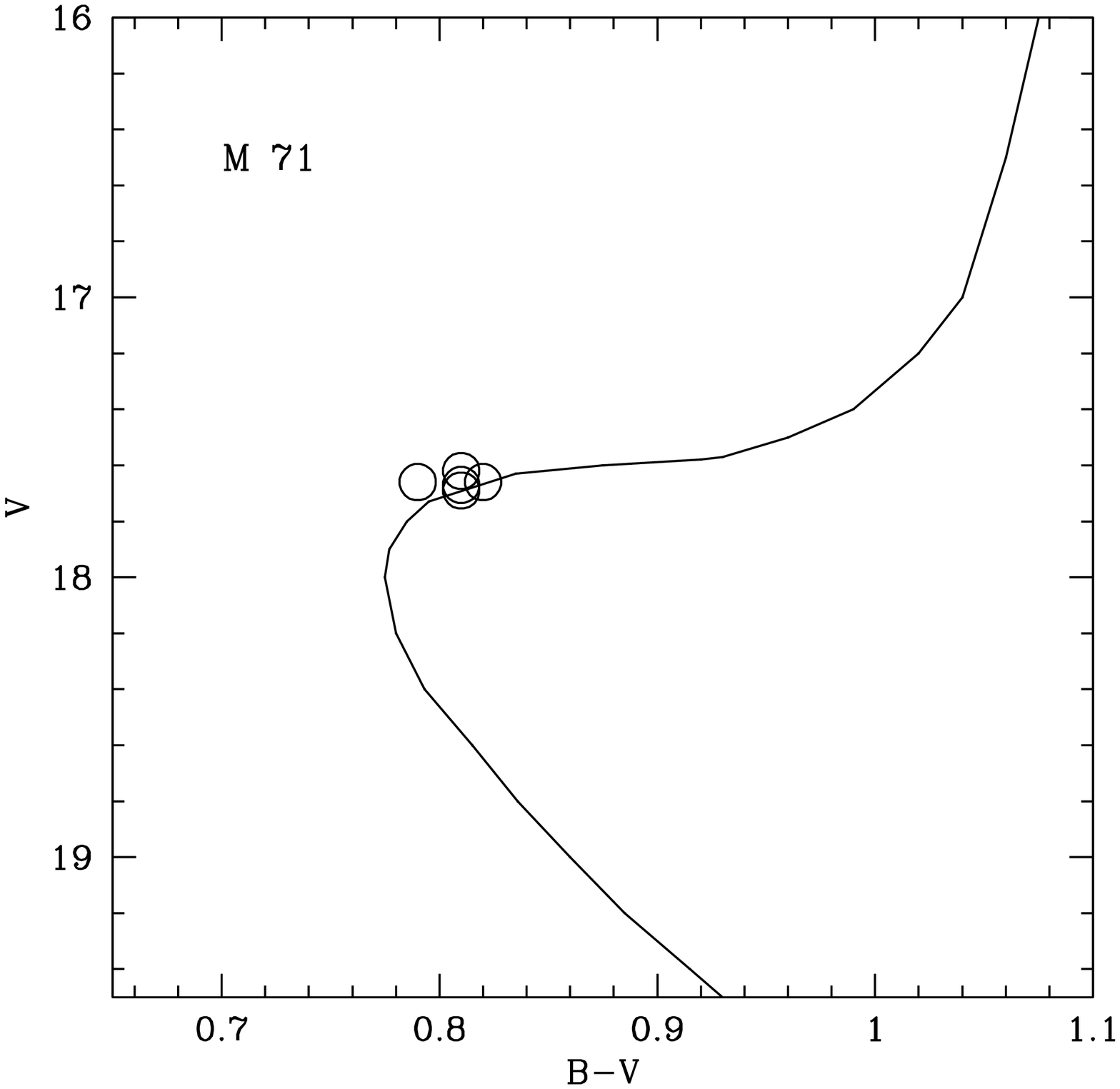}
\caption{The positions of our stars in the color-magnitude diagram.  The
fiducial sequence is from Hodder et al.~(1992).  The stars have nearly
identical values of V and B-V so have virtually the same temperatures and
luminosities as well as masses and ages.}
\end{figure}

\begin{figure}
\epsscale{0.8}
\plotone{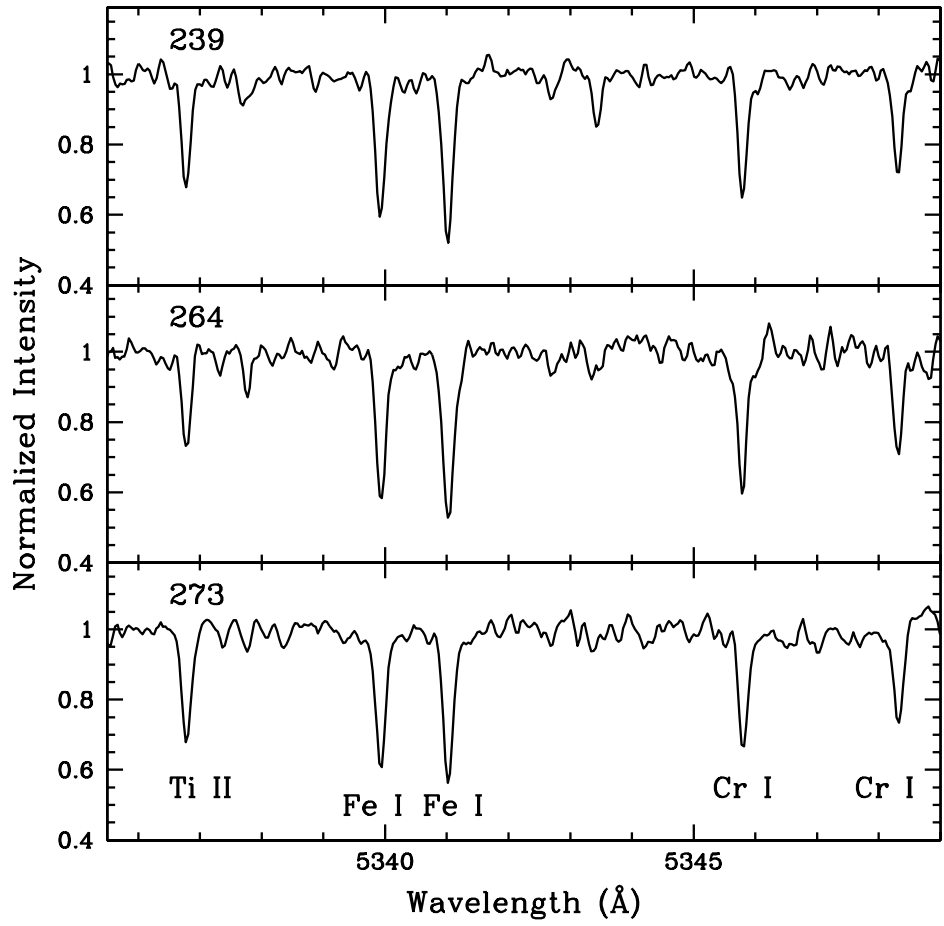}
%\epsscale{0.8}
\caption{Examples of part of one order of the spectra for three of our M71
stars.  This sample shows lines due to Fe I, Ti I and Cr I.\label{fig2}}
\end{figure}

\begin{figure}
\epsscale{0.8}
\plotone{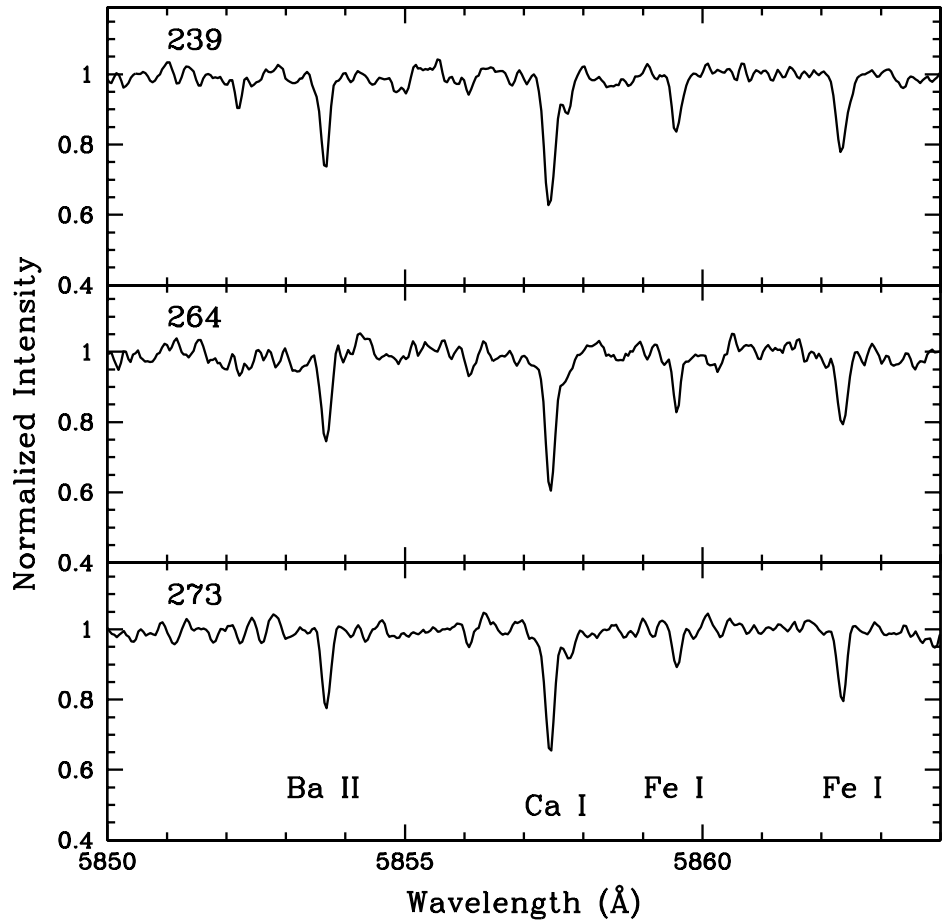}
\caption{Examples of part of another order of the spectra for three of our M71
stars.  This sample shows lines due to Fe I, Ba II and Ca I.\label{fig3}}
\end{figure}

\begin{figure}
\epsscale{1.0}
\plotone{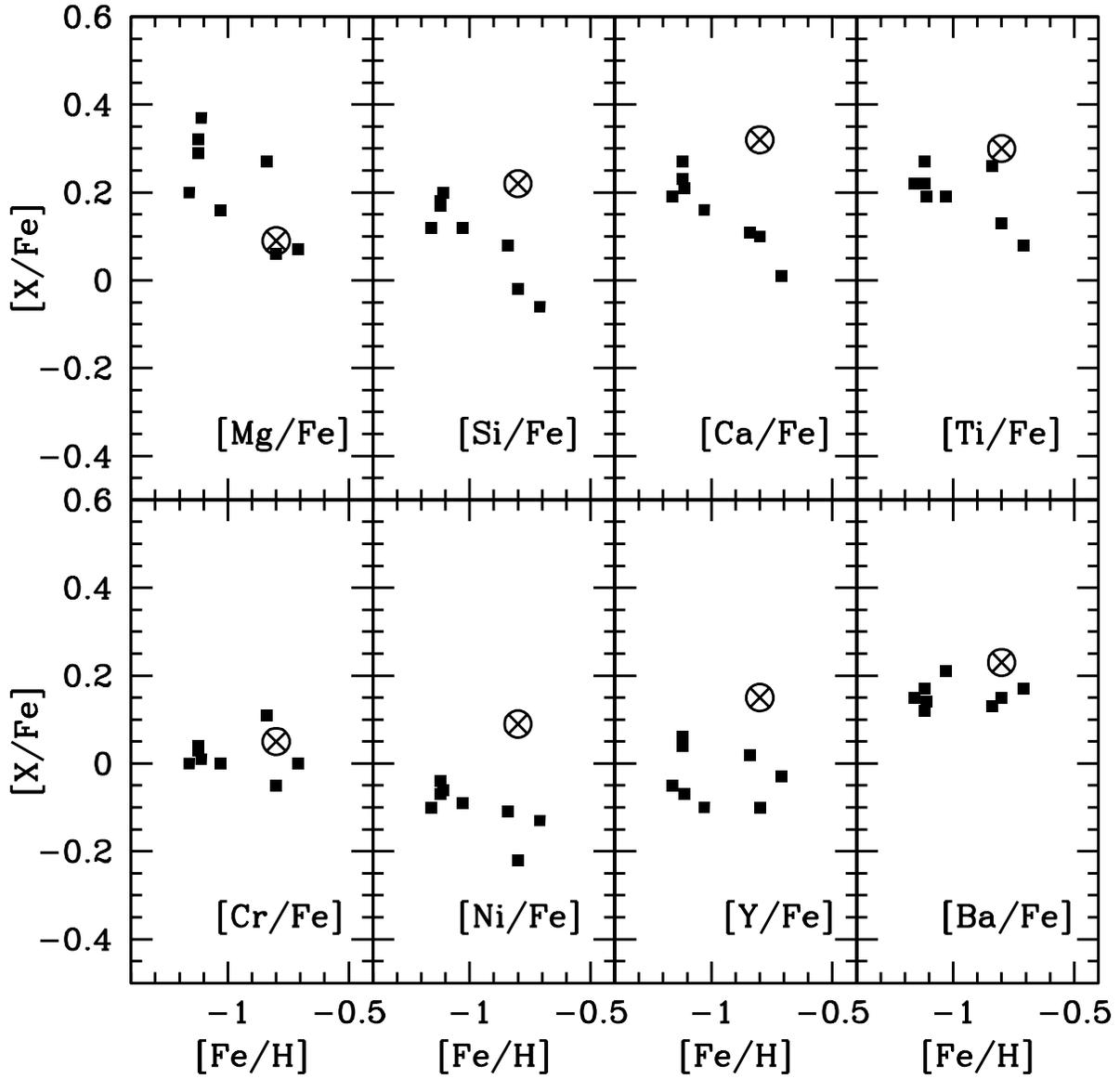}
\caption{Mean Fe-normalized M71 abundances(circled Xs) compared to halo field 
stars of similar metallicity from SB02 (filled squares).
}
\end{figure}

\begin{figure}
\plotone{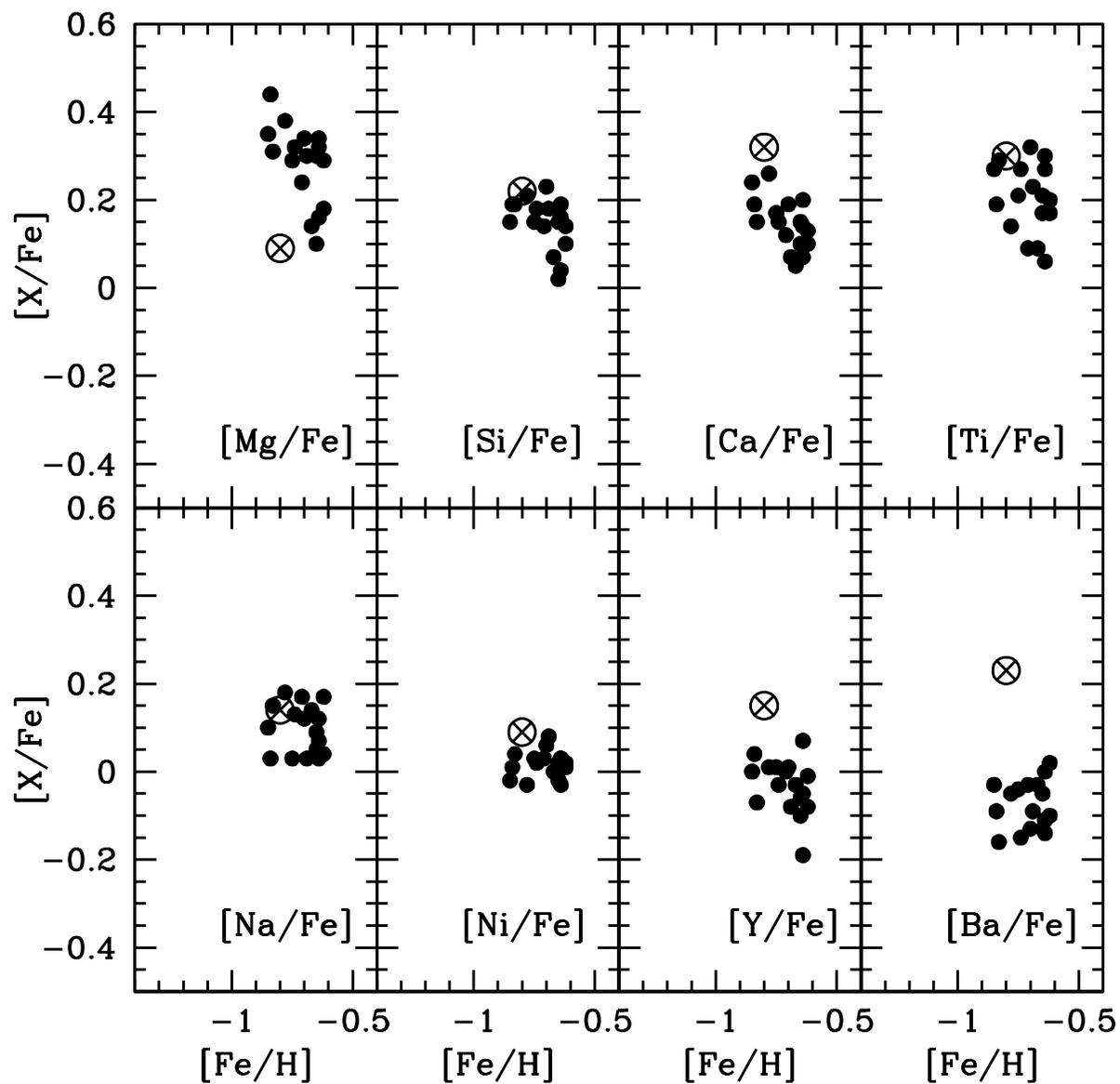}
\caption{ Mean Fe-normalized M71 abundances(circled Xs) compared to disk field
stars of similar metallicity from EAGLNT (filled circles).  This is similar to
Figure 4 except that the lower left panel is for [Na/Fe] rather than [Cr/Fe]
because EAGLNT did not determine Cr abundances.
}
\end{figure}

\begin{figure}
\plotone{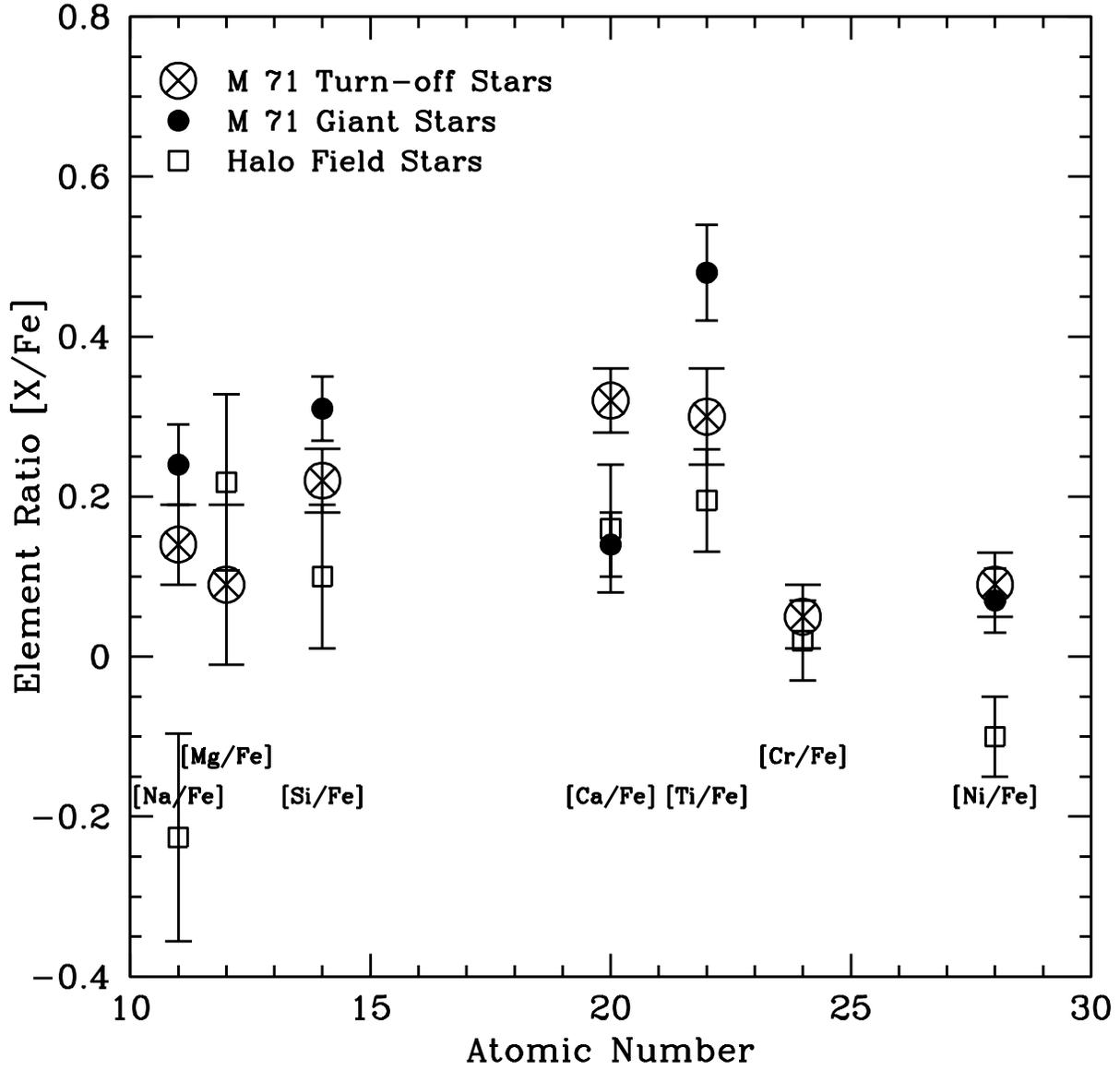}
\caption{Abundance comparisons of our mean abundances for the five M71 
turn-off stars with those of SKPLS for ten red giants in M71 and those of SB02
for eight field halo stars of similar [Fe/H] to M71.
}
\end{figure}

\begin{figure}
\plotone{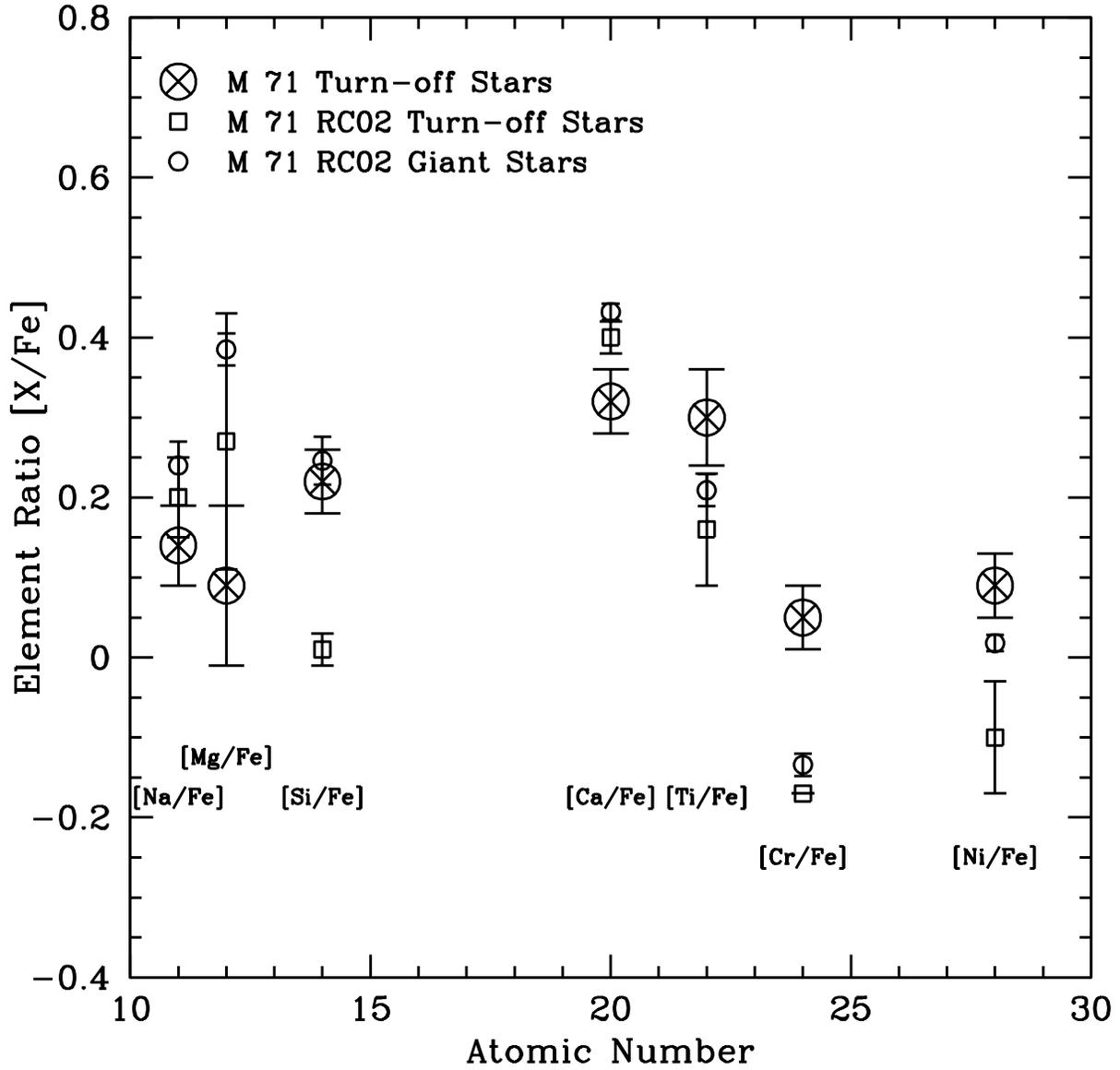}
\caption{Abundance comparisons of our mean abundances for the five M71 
turn-off stars with those of RC02 for three turn-off stars and for 19 red 
giants in M71.
}
\end{figure}

\begin{figure}
\plotone{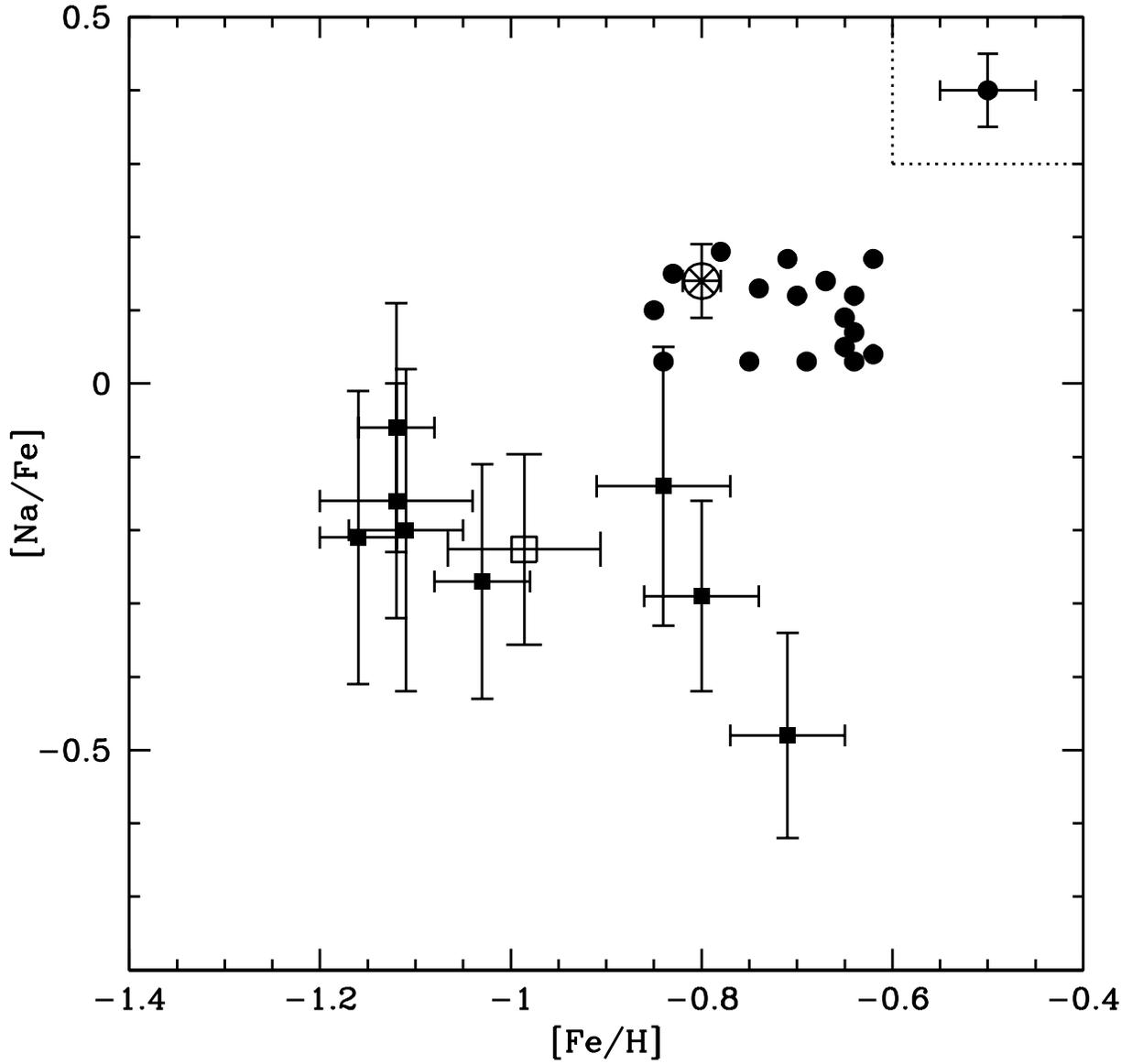}
\caption{Comparisons of [Na/Fe] for the mean of our five M71 turn-off stars
with those of EAGLNT for 16 disk stars (filled circles) and of SB02 for 
eight halo stars (filled squares).  The mean for the halo stars is given by
the open square.  Error bars are shown for M71 and the halo stars, but a
typical error bar for the disk stars is shown in the upper right corner.
}
\end{figure}

\begin{figure}
\plottwo{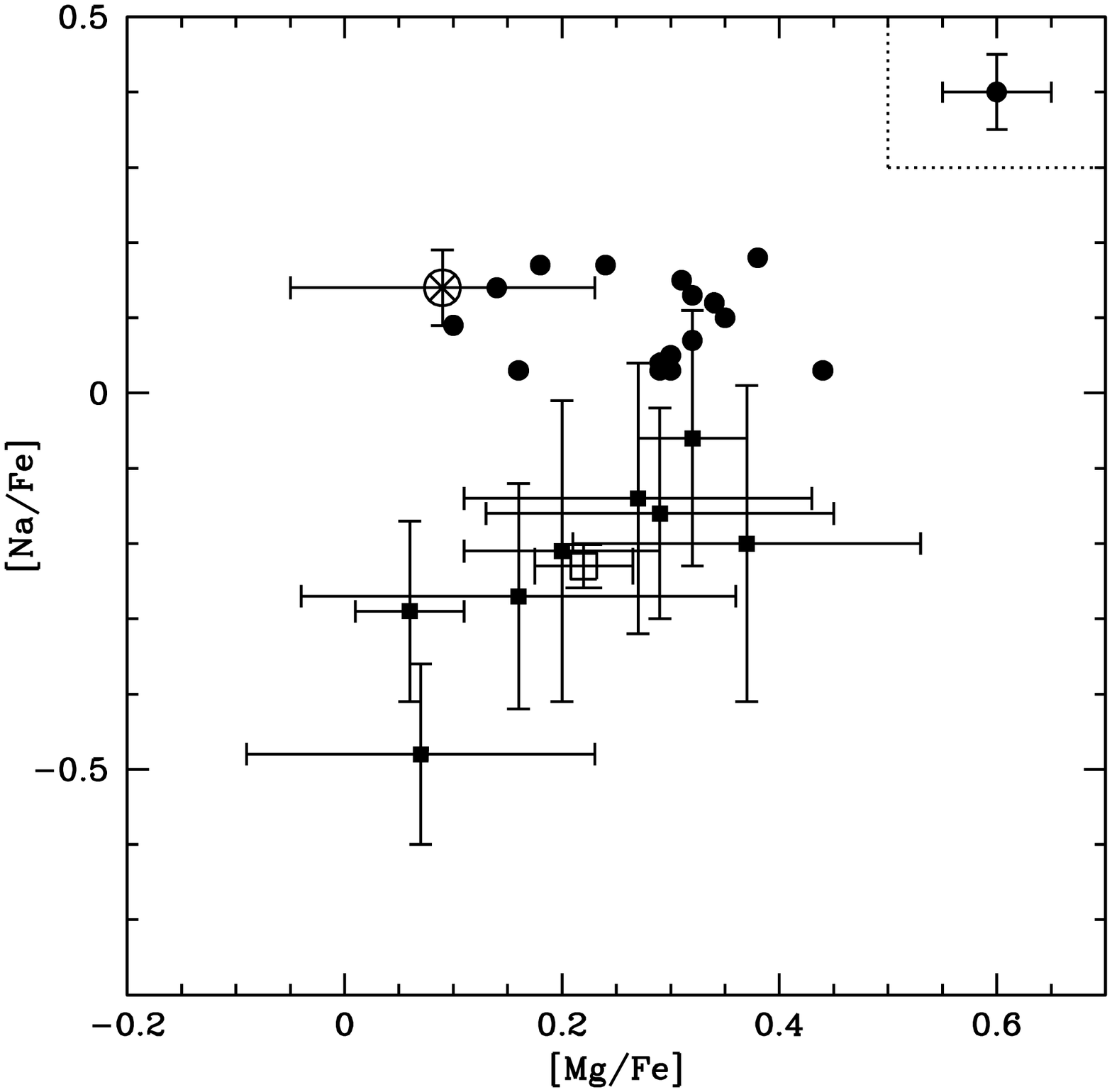}{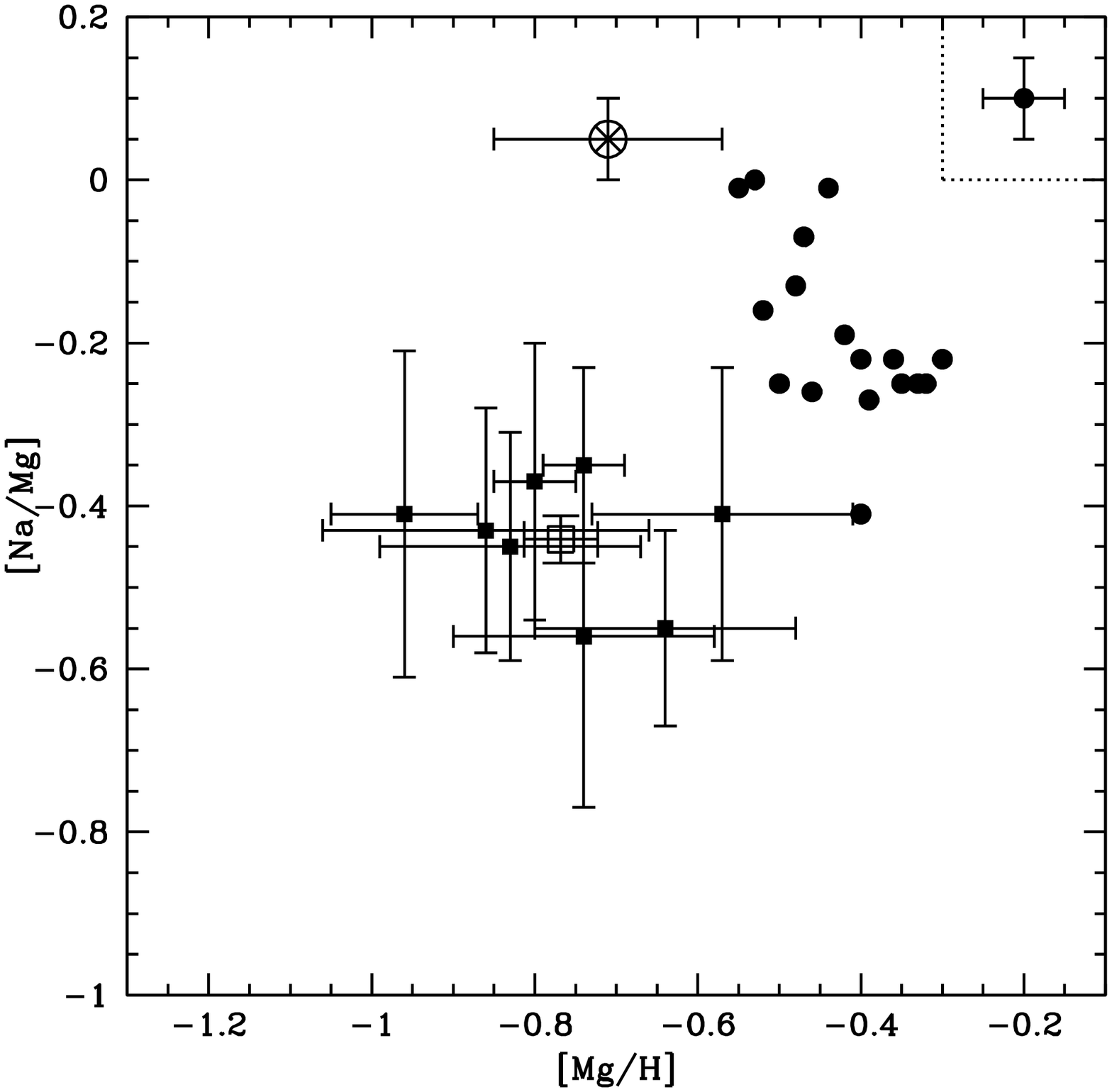}
\caption{Na and Mg abundances.  Left panel: [Na/Fe] vs [Mg/Fe] for M71
(circled X), the halo stars of SB02 (filled squares; open square represents
the mean) and disk stars of EAGLNT (filled circles).  In this comparison M71
appears to be high in Na and low in Mg.  Right panel: The [Na/Mg] vs [Mg/H]
plot where Mg replaces Fe in Figure 8.  M71 has a higher Na/Mg ratio than the
disk or halo field stars, but this may simply be the extension of the trend
toward higher [Na/Mg] with decreasing [Mg/H] evinced by the disk field stars.  
}
\end{figure}

%\begin{figure}
%\plotone{fig1.ps}
%\caption{
%}
%\end{figure}

\clearpage
%\input{tab1.tex}
%\documentclass[12pt,preprint]{aastex}
%\documentclass{aastex}
%\begin{document}
%% Note that the table will print double-spaced since we are using the
%% manuscript style. Change the style to preprint or preprint2 to see 
%% how LaTeX formats the table in those styles.

%\singlespace
\begin{center}
\begin{deluxetable}{lccclcc} 
\tablewidth{0pc}
\tablenum{1}
\tablecolumns{7} 
\tablecaption{M71 Observations} 
\tablehead{ 
\colhead{Star}  &  \colhead{V}  & \colhead{B-V}  & \colhead{(B-V)$_{\rm o}$} &
\colhead{Night\tablenotemark{1}} & \colhead{Total Exp. Time\tablenotemark{2}} 
& \colhead{Total}  \\
\colhead{} & \colhead{} & \colhead{} & \colhead{} & \colhead{} & \colhead{(hr min)} & \colhead{S/N} }
\startdata 
239..... & 17.62 & 0.81 & 0.53 & D, D, F$^3$, G  & 3 15 & 60  \\
259..... & 17.66 & 0.82 & 0.54 & A$^3$, B, C, C, G&4 00  & 50 \\
260..... & 17.66 & 0.79 & 0.51 & D, D, G$^3$    & 2 30  & 50 \\
264..... & 17.67 & 0.81 & 0.53 & D, D, D & 2 15 & 48 \\
273..... & 17.68 & 0.81 & 0.53 & B, C, C, E, F$^3$, F& 4 45 & 60\\
\enddata 
\tablenotetext{1}{A=1996 July 26; B=1996 July 27; C=1997 August 30; 
D=1997 August 31; E=1998 June 23;  
F=1998 September 10; G=1998 September 11.}
\tablenotetext{2}{Individual Exposures are 45 minutes, unless otherwise noted.}
\tablenotetext{3}{60-minute exposure.}

\end{deluxetable} 
\end{center}

%\end{document}

%% 
%% End of file `obs_table.tex'. 

\clearpage
%\input{tab2.tex}
%\documentclass[12pt,preprint]{aastex}
%\documentstyle[12pt,aasms4]{article}
%\documentstyle[apjpt4]{article}

%\begin{document}
%\tightenlines

%\setcounter{page}{0}
%\ptlandscape
%\vspace*{1.5in}
%\hspace*{0.25in}

\begin{deluxetable} {llrrrrrrr}
\small
\tablecolumns{9}
\tablewidth{0pc}
\tablenum{2}
\tablecaption{Measured Equivalent Widths}
\tablehead{
\multicolumn{1}{c}{$\lambda$ } &
\multicolumn{1}{c}{Ex. Pot.} &
\multicolumn{1}{c}{$\log{gf}$} &
\multicolumn{5}{c}{Equivalent Widths}\\
\multicolumn{1}{c}{({\AA})} &
\multicolumn{1}{c}{(eV)} &
\multicolumn{1}{c}{} &
\multicolumn{1}{c}{239} &
\multicolumn{1}{c}{259} &
\multicolumn{1}{c}{260} & 
\multicolumn{1}{c}{264} &
\multicolumn{1}{c}{273} &
\multicolumn{1}{c}{Sun} 
}
\startdata
\multicolumn{1}{l}{\ion{Na}{1}} &
\multicolumn{8}{c}{} \\
\hline
%\ion{Na}{1}      &	   &	     &	    &	   &	  &	 &	&    \\
%\cline{1-9}\\
5682.650 & 2.1000  & $-$0.8200 & 40.2 & 45.6 & 48.1 & 38.6 & 40.1 & 106.5 \\
5688.219 & 2.1000  & $-$0.3700 & 49.5 & 69.9 & 63.7 & 65.4 & 62.2 & \nodata \\
6154.227 & 2.1000  & $-$1.6600 &  7.4 & 13.9 & 14.7 & \nodata &  8.6 & 37.8 \\
6160.751 & 2.1000  & $-$1.3500 & 12.5 & \nodata & 16.4 & 27.3 & 15.5 & 57.5 \\

%\cline{1-9} \\
% \ion{Mg}{1}      &	   &	     &	    &	   &	  &	 &	&    \\
%\cutinhead{\ion{Mg}{1} \phm{blahhhhhhhhhhhhhhhhhhhhhhhhhhhhhhhhhhhhhhhhhhhhhhhhhhhhhhhhh}}
%\cline{1-9}\\
\hline
\multicolumn{1}{l}{\ion{Mg}{1}} &
\multicolumn{8}{c}{} \\
\hline
4571.099 & 0.000  & $-$5.5850 & 71.5  & 71.5  & 73.3  & 63.0  & 71.3 & 111.0 \\
4703.98  & 4.340  & $-$0.5350  & 160.3 & 148.6 & 118.5 & 149.8 & 132.9 & \nodata \\
4730.028 & 4.340  & $-$2.4300 & 17.7  & 14.7  & \nodata & \nodata & 17.5 & 74.0\\
6318.72  & 5.110  & $-$1.6840 & \nodata & \nodata & \nodata & \nodata & \nodata
& 50.8 \\

%\cline{1-9} \\
%\cutinhead{\ion{Si}{1} \phm{blahhhhhhhhhhhhhhhhhhhhhhhhhhhhhhhhhhhhhhhhhhhhhhhhhhhhhhhhh}}
%\cline{1-9}\\
\hline
\multicolumn{1}{l}{\ion{Si}{1}} &
\multicolumn{8}{c}{} \\
\hline
5772.148 & 5.0809  & $-$1.7500 & 19.6 & 30.2 & 30.2 & 33.2 & 15.8 & 53.0 \\
5948.545 & 5.0813  & $-$1.2250 & 54.9 & 40.0 & 12.9 & 42.5 & 49.0 & 84.1 \\
6155.141 & 5.6200  & $-$0.8400 & 44.0 & 43.9 & 37.1 & 47.0 & 41.7 & 83.4 \\

%\cline{1-9} \\
%\cutinhead{\ion{Si}{2} \phm{blahhhhhhhhhhhhhhhhhhhhhhhhhhhhhhhhhhhhhhhhhhhhhhhhhhhhhhhhh}}
%\cline{1-9}\\
\hline
\multicolumn{1}{l}{\ion{Si}{2}} &
\multicolumn{8}{c}{} \\
\hline
6347.100 & 8.1200  &   0.3199 & 28.3 & 23.0 & 23.0 & 29.9 & 30.7 & 45.5 \\

%\cline{1-9} \\
%\cutinhead{\ion{Ca}{1} \phm{blahhhhhhhhhhhhhhhhhhhhhhhhhhhhhhhhhhhhhhhhhhhhhhhhhhhhhhhhh}}
%\cline{1-9}\\
\hline
\multicolumn{1}{l}{\ion{Ca}{1}} &
\multicolumn{8}{c}{} \\
\hline
4526.934 & 2.7092  & $-$0.4890 & 50.0  & 53.1  & 42.8  & 56.8 & 44.5 & \nodata \\
4578.551 & 2.5214  & $-$0.6285 & 49.3  & 40.7  & \nodata & 51.6 & 49.6 & \nodata \\
4685.268 & 2.9327  & $-$0.8800 & 32.2  & 12.1  & 23.5  & \nodata & 14.8 & \nodata \\
5188.844 & 2.9327  & $-$0.0900 & \nodata & \nodata & \nodata & \nodata & \nodata & 91.1 \\  
5260.389 & 2.5210  & $-$1.8095 & 13.9 & 11.1 & 9.1 & 10.9 & \nodata & \nodata \\
5261.707 & 2.5214  & $-$0.6545 & 62.6  & 58.0  & 45.5  & 60.9 & 62.1 & 98.2 \\
5512.980 & 2.9327  & $-$0.3685 & 49.7  & 46.5  & 47.3  & 34.9 & 51.8 & 86.5 \\
5581.968 & 2.5232  & $-$0.6325 & 73.2  & 55.2  & 50.3  & 66.3 & 66.3 & 94.4 \\
%5588.755 & 2.5259  &  0.2840 & \nodata & \nodata & \nodata & \nodata & \nodata & \nodata \\
5590.117 & 2.5214  & $-$0.6405 & 56.1  & 52.1  & 63.1  & 63.3 & 50.8 & 91.1 \\
5598.480 & 2.5214  & $-$0.2200 & \nodata & \nodata & 98.6  & \nodata & \nodata &
138.4 \\ 
5601.277 & 2.5259  & $-$0.6900 & 69.3  & 55.8  & \nodata & 60.5 & 56.2 & \nodata
\\ 
5857.451 & 2.9327  &   0.2350 & 86.7  & 80.8  & 83.0  & 84.5 & 69.1 & 138.4 \\
6163.755 & 2.5214  & $-$1.2860 & 31.9  & 25.5 & 27.3  & 33.0  & 23.7 & \nodata \\
6166.440 & 2.5214  & $-$1.1400 & 34.6  & 37.6 & 25.3  & 36.6  & 23.9 & 70.3 \\
6169.042 & 2.5232  & $-$0.7970 & 59.3  & 48.0 & 49.7  & 58.9  & 49.0 & 93.7 \\
6169.562 & 2.5259  & $-$0.3740 & 60.6  & 56.4 & \nodata & 67.8  & 61.1 & 112.6 \\
6449.810 & 2.5214  & $-$0.5020 & 67.0  & 68.7 & 50.5  & 65.3  & 52.3 & 107.8 \\
6455.600 & 2.5230  & $-$1.3400 & 30.0 & 23.2 & 28.1 & 24.9  & 30.0 & 56.0 \\
6471.661 & 2.5259  & $-$0.6380 & 59.2  & \nodata & 41.7  & 56.4  & \nodata & 92.6 \\
6493.781 & 2.5214  &   0.0155 & 87.4  & 87.2 & 75.5 & 98.5  & 76.8 & 126.7 \\
6499.650 & 2.5232  & $-$0.8180 & 44.0  & 38.4 & 33.0 & 57.6 & 36.1 & 86.3 \\
6717.686 & 2.709   & $-$0.5670 & 56.2  & 38.2 & 59.8 & 52.8 & 46.8 & \nodata \\

%\cline{1-9}
%\cutinhead{\ion{Ti}{1} \phm{blahhhhhhhhhhhhhhhhhhhhhhhhhhhhhhhhhhhhhhhhhhhhhhhhhhhhhhhhh}}
%\cline{1-9}\\
\hline
\multicolumn{1}{l}{\ion{Ti}{1}} &
\multicolumn{8}{c}{} \\
\hline
4518.023 & 0.8259  & $-$0.2690 & 48.0 & 39.7 & 38.3 & 40.5 & 31.7 & 69.8 \\
4527.305 & 0.8130  & $-$0.4700 & 42.2 & 40.4 & 40.9 & 46.4 & 38.2 & \nodata \\
%4533.239 & 0.8484  &  0.5320 & \nodata & \nodata & \nodata & \nodata & \nodata & \nodata \\
4534.778 & 0.8360  &  0.3360 & \nodata &\nodata  & 71.1 & \nodata & \nodata & \nodata\\
4535.570 & 0.8259  &  0.1200 & 67.7 & \nodata & \nodata & 69.0 & 63.0 & \nodata \\
4544.688 & 0.8182  & $-$0.5200 & \nodata & 35.2 & 39.9 & \nodata & \nodata & \nodata \\
4617.254 & 1.7489  &  0.4450 & 25.6 & 28.1 & \nodata & 37.2 & 33.5 & 60.6 \\
4681.908 & 0.0480  & $-$1.0150 & 58.2 & 38.4 & \nodata & 51.7 & 33.0 & \nodata \\
4840.874 & 0.8996  & $-$0.4530 & 35.2 & 40.4 & 32.8 & 52.1 & 24.5 & 66.4 \\
%4981.732 & 0.8484  &  0.5600 & \nodata & \nodata & \nodata & \nodata &
%\nodata  & \nodata \\
%4991.067 & 0.8360  &  0.4360 & \nodata & \nodata & \nodata & \nodata & \nodata& \nodata \nl 
4999.504 & 0.8259  &  0.3060 & 74.5 & 66.7 & \nodata & \nodata & 74.1 & \nodata \\
5016.162 & 0.8484  & $-$0.5180 & 42.4 & 28.1 & 35.0 & 29.6 & 32.2 & 64.7 \\
5020.028 & 0.8360  & $-$0.3580 & 49.1 & 51.8 & 36.5 & 52.4 & 45.9 & 78.4 \\
5022.871 & 0.8259  & $-$0.3780 & 54.5 & 49.3 & 45.8 & 48.3 & 35.2 & 70.1 \\
5035.907 & 1.4602  &  0.2600 & \nodata & \nodata & 59.0 & \nodata & \nodata & \nodata \\
5036.468 & 1.4432  &  0.1860 & 43.7 & 43.2 & 53.4 & 51.5 & 47.8 & 74.3 \\
5039.959 & 0.0211  & $-$1.1300 & 49.6 & 49.0 & 34.2 & 50.5 & 42.4 & 73.2 \\
5064.654 & 0.0480  & $-$0.8550 & 58.8 & 46.1 & 41.3 & 62.9 & 56.0 & \nodata \\
5192.969 & 0.0211  & $-$0.9500 & 66.1 & 54.0 & 48.0 & 51.6 & \nodata & \nodata \\

%\cline{1-9}
%\cutinhead{\ion{Ti}{2} \phm{blahhhhhhhhhhhhhhhhhhhhhhhhhhhhhhhhhhhhhhhhhhhhhhhhhhhhhhhhh}}
%\cline{1-9}\\
\hline
\multicolumn{1}{l}{\ion{Ti}{2}} &
\multicolumn{8}{c}{} \\
\hline
4501.272 & 1.1156  & $-$0.7550 & \nodata & \nodata & \nodata  & 112.7 & \nodata & \nodata \\
4563.761 & 1.2214  & $-$0.9600 & \nodata & \nodata  & \nodata & 103.0 & \nodata & \nodata \\
4571.968 & 1.5719  & $-$0.5300 & \nodata & \nodata & \nodata & 108.2 & \nodata & \nodata \\
4589.958 & 1.2369  & $-$1.7900 & 61.6 & 70.5 & 47.1 & 73.9  & 54.0 & \nodata \\
4657.203 & 1.2430  & $-$2.2350 & \nodata & 36.9 & 27.1 & 43.9  & 43.7 & 51.4 \\
4779.985 & 2.0478  & $-$1.3700 & 50.0 & 53.3 & 48.0 & 56.5  & 47.4 & 62.2 \\
4805.085 & 2.0614  & $-$1.1000 & 63.7 & 66.4 & 56.6 & 80.0  & 58.1 & \nodata \\
5129.152 & 1.8918  & $-$1.3900 & \nodata & 48.5  & 69.3  & 73.6  & 45.8 & 79.4 \\
5154.070 & 1.5659  & $-$1.9200 & 53.7  & 44.8  & 43.2  & 75.3  & 47.2 & 72.4 \\
%5188.680 & 1.5819  & $-$1.2100 & \nodata & \nodata & \nodata & \nodata & \nodata & \nodata \nl 
5226.543 & 1.5659  & $-$1.3000 & 78.6  & 81.9  & 60.9  & 82.9  & 73.8 & \nodata \\
5336.781 & 1.5819  & $-$1.6650 & 59.4  & 50.0  & 54.7  & 49.6  & 61.4 & 71.4 \\
5381.018 & 1.5658  & $-$2.0250 & 48.9  & 45.4  & 48.4  & 66.9  & 46.8 & 59.6 \\

%\cline{1-9}
%\cutinhead{\ion{Cr}{1} \phm{blahhhhhhhhhhhhhhhhhhhhhhhhhhhhhhhhhhhhhhhhhhhhhhhhhhhhhhhhh}}
%\cline{1-9}\\
\hline
\multicolumn{1}{l}{\ion{Cr}{1}} &
\multicolumn{8}{c}{} \\
\hline
4496.842 & 0.9415  & $-$1.1500 & \nodata & \nodata & 67.3 & \nodata & \nodata & \nodata \\
4540.734 & 3.1046  &  0.0280 & \nodata & 16.6  & \nodata & \nodata & \nodata & 84.5 \\  
4545.945 & 0.9415  & $-$1.3700 & 43.7 & \nodata &\nodata & \nodata & 41.5 & 84.5 \\
4591.389 & 0.9685  & $-$1.7400 & 26.1 & \nodata & \nodata & 44.9  & \nodata & 82.6 \\
4600.741 & 1.0037  & $-$1.2600 & 46.7 & 69.3  & \nodata & 61.6  & 42.7 & 92.1 \\
4616.120 & 0.9829  & $-$1.1900 & 42.7 & 55.4  & 44.1 & 55.5  & 45.8 & 87.6 \\
4626.174 & 0.9685  & $-$1.3200 & 48.0 & 44.4  & 49.3 & 44.0  & 38.9 & 78.7 \\
4646.148 & 1.0301  & $-$0.7000 & 68.9 & 66.5  & \nodata & 67.9  & \nodata & \nodata \\
4651.282 & 0.9829  & $-$1.4600 & 41.5 & 44.8  & 44.8 & 58.9  & 31.4 & 77.0 \\
4652.152 & 1.0037  & $-$1.0300 & 55.4 & 51.8  & 42.7 & 64.5  & 48.8 & 99.2 \\
4718.426 & 3.1955  &  0.0900 & \nodata & 23.3 & \nodata & 21.6  & \nodata & 65.4 \\  
%5206.038 & 0.9415  &  0.0190 & \nodata & \nodata & \nodata & \nodata  & \nodata & \nodata \nl
%5208.419 & 0.9415  &  0.1580 & \nodata & \nodata & \nodata & \nodata & \nodata & \nodata \nl 
5247.566 & 0.9610  & $-$1.6400 & 32.0 & 31.9 & 32.3 & 41.8  & 36.6 & 81.5 \\
5296.691 & 0.9829  & $-$1.4000 & 53.4 & 42.1 & 47.0 & 62.1  & 46.5 & 91.9 \\
5298.277 & 0.9829  & $-$1.1600 & \nodata & 53.7  & 63.7 & 79.7  & 57.3 & \nodata \\
5345.801 & 1.0037  & $-$0.9800 & 62.1 & 70.4  & 57.5 & 70.8  & 64.3 & 116.0 \\
5348.312 & 1.0037  & $-$1.2900 & 49.8 & 52.8  & 41.2 & 52.9  & 51.3 & 100.5 \\
5409.772 & 1.0301  & $-$0.7150 & 69.3 & 85.1  & 73.5 & \nodata & 70.4 & 141.4 \\
6330.093 & 0.9415  & $-$2.9200 & \nodata & \nodata  & \nodata & \nodata & \nodata & 27.3 \\

%\cline{1-9}
%\cutinhead{\ion{Cr}{2} \phm{blahhhhhhhhhhhhhhhhhhhhhhhhhhhhhhhhhhhhhhhhhhhhhhhhhhh}}
%\cline{1-9}\\
\hline
\multicolumn{1}{l}{\ion{Cr}{2}} &
\multicolumn{8}{c}{} \\
\hline
4558.650 & 4.0737  & $-$0.6600 & 47.2 & 48.5 & 44.8 & 51.7 & 45.0 & 78.8 \\
4588.199 & 4.0715  & $-$0.6300 & 44.4 & 49.3 & 42.0 & 39.9 & 41.5 & 67.9 \\
4634.070 & 4.0725  & $-$1.2400 & \nodata & 21.9 & 26.2 & 33.3 & 36.9 & 61.5 \\
4848.235 & 3.8647  & $-$1.1400 & 30.5 & 40.7 & 20.1 & \nodata & 42.6 & 65.6 \\
5237.329 & 4.0737  & $-$1.1600 & 28.3 & 31.9 & 33.0 & 28.4 & 27.7 & 51.5 \\

\tablebreak
%\cline{1-9}
%\cutinhead{\ion{Fe}{1} \phm{blahhhhhhhhhhhhhhhhhhhhhhhhhhhhhhhhhhhhhhhhhhhhhhhhhhh}}
%\cline{1-9}\\
%\hline
\multicolumn{1}{l}{\ion{Fe}{1}} &
\multicolumn{8}{c}{} \\
\hline
4489.739 & 0.1210  & $-$3.9318 & 60.4  & 57.9  & 50.6  & 54.8  & 61.5 & \nodata \\
%4494.563 & 2.1980  & $-$1.1395 & \nodata & \nodata & \nodata  & \nodata & \nodata & \nodata \nl
%4528.614 & 2.1760  & $-$0.8870 & \nodata & \nodata & \nodata & \nodata &  \nodata &\nodata \nl
%4531.148 & 1.4850  & $-$2.1280 & \nodata & \nodata & \nodata & \nodata &  \nodata & \nodata \nl
4556.126 & 3.6030  & $-$0.7870 & 61.9  & 67.8  & 47.0 & 61.2  & 53.3 & \nodata \\
4592.651 & 1.5580  & $-$2.4556 & 66.3  & \nodata & \nodata  & \nodata &  \nodata & \nodata \\
4602.000 & 1.6080  & $-$3.1439 & 24.2  & 27.7  & \nodata & \nodata &  \nodata & \nodata \\
4602.941 & 1.4850  & $-$2.2080 & \nodata & \nodata & \nodata & \nodata &  \nodata & 67.6 \\
4619.287 & 3.6030  & $-$1.1200 & 41.4  & 45.8 & 38.4  & 46.9  & 49.4 & \nodata \\
4625.044 & 3.2410  & $-$1.3400 & 45.7  & 55.6  & 50.7  & 55.6  & 48.2 & \nodata \\
4630.120 & 2.2790  & $-$2.5935 & 42.0  & \nodata & 35.2  & 44.3  & 34.1 & \nodata \\
4637.503 & 3.2830  & $-$1.3900 & 54.4  & 38.1  & 35.7  & 44.2  & 48.1 & \nodata \\
4638.010 & 3.6030  & $-$1.1195 & \nodata  & 33.0  & 48.7  & 31.6  & 42.7 & \nodata \\
4647.434 & 2.9490  & $-$1.3305 & 60.7  & 62.7  & 46.9  & 50.6  & 56.9 & \nodata \\
4669.171 & 3.6540  & $-$1.3100 & \nodata   & 33.5  & 22.1  & 24.1  & 36.6 & \nodata \\
4678.846 & 3.6030  & $-$0.7465 & 53.1  & 53.6  & 45.6  & 57.8  & 45.4 & \nodata \\
4691.411 & 2.9910  & $-$1.4865 & 55.7  & \nodata & 51.7  & 53.9  & 62.6 & \nodata \\
4710.283 & 3.0180  & $-$1.6120 & 56.6  & 53.9  & 47.0  & 60.4  & 50.0 & \nodata \\
4728.546 & 3.6540  & $-$1.1710 & 38.2  & 27.6  & 39.2  & 51.5  & 27.8 & \nodata \\
4733.591 & 1.4850  & $-$2.9872 & 47.3  & 46.1  & 59.8  & 54.8  & 50.1 & \nodata \\
4736.773 & 3.2110  & $-$0.7460 & 67.9  & \nodata & 62.0  & 70.2  & 69.9 & \nodata \\
4741.529 & 2.8320  & $-$1.8820 & 28.8  & 30.8  & 30.0  & 43.1  & 31.5 & 71.4 \\
%4871.318 & 2.8660  & $-$0.3860 & \nodata & \nodata & \nodata  & \nodata & \nodata & \nodata \nl
%4872.137 & 2.8820  & $-$0.5835 & \nodata & \nodata & \nodata & \nodata & \nodata & \nodata \nl
4882.144 & 3.4170  & $-$1.6400 & 35.1  & 32.7  & \nodata & \nodata & 27.3 & 73.6 \\
%4918.994 & 2.8660  & $-$0.3560 & \nodata & \nodata & \nodata  & \nodata &  \nodata & \nodata \nl
%4920.503 & 2.8330  &  0.0640 & \nodata & \nodata & \nodata  & \nodata & \nodata & \nodata \nl
4924.770 & 2.2790  & $-$2.2305 & 38.5  & 53.9  & 31.6  & 35.1  & 37.7 & \nodata \\
4938.814 & 2.8760  & $-$1.0770 & 73.8  &\nodata  & 66.3  & \nodata  & 64.6 & \nodata \\
4939.687 & 0.8590  & $-$3.3152 & \nodata  & 55.4  & 50.9  & 72.4  & 60.3 & \nodata \\
4946.385 & 3.3690  & $-$1.1700 & 58.9  & 51.0  & 45.2  & 68.9  & 53.6 & \nodata \\
4985.253 & 3.9290  & $-$0.5590 & 42.8  & 44.8  & 41.1  & 43.6  & 59.1 & \nodata \\
4985.547 & 2.8650  & $-$1.3310 & 59.7  & 58.7  & 46.6  & 57.5  & 50.9 & \nodata \\
4994.130 & 0.9150  & $-$2.9690 & 68.0  & \nodata & 59.1  & \nodata & 68.3 & \nodata \\
5001.862 & 3.8820  &  0.0100 & \nodata & \nodata & \nodata & \nodata & 70.5 & \nodata \\
%5006.119 & 2.8330  & $-$0.6385 & \nodata & \nodata & \nodata & \nodata &  \nodata & \nodata \nl
%5012.068 & 0.8590  & $-$2.6231 & \nodata & \nodata  & \nodata & \nodata &  \nodata & \nodata \nl
5022.236 & 3.9840  & $-$0.5300 & \nodata & 52.7 & 50.6  & 70.6  & 57.1 & \nodata \\
5028.127 & 3.5730  & $-$1.1220 & 47.5  & 41.6 & 39.1  & \nodata  & 48.5 & \nodata \\
5044.212 & 2.8510  & $-$2.0590 & 38.4  & 34.8 & 23.6  & 35.8  & 30.9 & 70.7 \\
5049.819 & 2.2790  & $-$1.3495 & \nodata & \nodata & 78.1  & \nodata & \nodata & \nodata \\
5051.635 & 0.9150  & $-$2.7640 & \nodata & \nodata & 76.6  & \nodata & \nodata & \nodata \\
5068.766 & 2.9400  & $-$1.1355 & 70.6  & \nodata  & 64.1  &  \nodata & 72.5 & \nodata \\
5079.224 & 2.1980  & $-$2.0860 & 70.6  & 63.2  & 52.2  & 57.8  & 61.8 & \nodata \\
5079.740 & 0.9900  & $-$3.2328 & 68.2  & 75.6  & 53.7  & \nodata  & 50.2 & \nodata \\
5083.339 & 0.9580  & $-$2.8419 & 74.2  & 73.1  & 65.4  & \nodata  & 65.0 & \nodata \\
%5098.697 & 2.1760  & $-$2.0260 & \nodata & \nodata & \nodata & \nodata  &  \nodata & \nodata \nl
5110.413 & 0.0000  & $-$3.7595 & \nodata & \nodata & 79.7 & \nodata &  \nodata & \nodata \\
5123.720 & 1.0110  & $-$3.0630 & 70.1  & 75.6  & \nodata & 67.8  & 62.3 & \nodata \\
5150.840 & 0.9900  & $-$3.0200 & \nodata & 73.6  & 54.3  & 62.1  & 63.2 & \nodata \\
5151.911 & 1.0110  & $-$3.3215 & 61.3  & 56.6  & 42.7  & 60.2  & 51.5 & \nodata \\
%5171.596 & 1.4850  & $-$1.7570 & \nodata  & \nodata & \nodata & \nodata &  \nodata & \nodata \nl
%5192.344 & 2.9980  & $-$0.4210 & \nodata & \nodata & \nodata & \nodata &  \nodata & \nodata \nl
%5194.942 & 1.5580  & $-$2.0555 & \nodata & \nodata & \nodata & \nodata &  \nodata & \nodata \nl
5198.711 & 2.2230  & $-$2.1130 & 56.8  & 49.4  & 48.6  & 48.8  & 55.4 & \nodata \\
5202.336 & 2.1760  & $-$1.8545 & \nodata & \nodata & 78.3  & \nodata & 78.9 & \nodata \\
5215.182 & 3.2660  & $-$0.8710 & 59.8  & 69.2  & 56.4  & 56.7  & 60.6 & \nodata \\
5216.274 & 1.6080  & $-$2.1160 & \nodata & \nodata  & 72.7  & \nodata &  \nodata & \nodata \\
5217.390 & 3.2110  & $-$1.1160 & 56.6  & 71.3  & 46.2  & 55.6  & 41.9 & \nodata \\
5225.525 & 0.1100  & $-$4.7696 & 38.0  & 37.1  & 40.3  & 41.3  & 34.5 & 73.5 \\
%5227.190 & 1.5570  & -1.2270 & \nodata & \nodata & \nodata & \nodata &  \nodata & \nodata \nl
%5232.940 & 2.9400  & -0.0960 & \nodata & \nodata & \nodata & \nodata &  \nodata & \nodata \nl
5242.491 & 3.6350  & $-$0.9035 & 49.7  & 49.9  & 40.4  & 56.4  & 42.6 \\
5250.210 & 0.1210  & $-$4.9172 & 32.7 & 12.5 & 22.6  & 32.3  & 28.0 & 68.9 \\
5250.646 & 2.1980  & $-$2.1150 & 61.8  & 62.5  & 53.5  & \nodata & 62.0 & \nodata \\
5263.305 & 3.2660  & $-$0.9245 & 64.1  & 74.0  & 64.6  & 56.4  & 60.1 & \nodata \\
%5269.537 & 0.8590  & -1.3225 & \nodata & \nodata & \nodata & \nodata &  \nodata & \nodata \nl
5307.361 & 1.6080  & $-$2.9496 & 52.3  & 53.4  & 45.9  & 47.1  & 50.2 & \nodata \\
%5328.039 & 0.9150  & -1.4655 & \nodata & \nodata & \nodata & \nodata &\nodata & \nodata \nl
%5328.532 & 1.5570  & -1.8500 & \nodata & \nodata & \nodata & \nodata & \nodata& \nodata \nl
5332.900 & 1.5570  & $-$2.8579 & 54.3  & 53.5  & 50.8  & 70.5  & 44.7 & \nodata \\
5339.930 & 3.2660  & $-$0.6800 & \nodata & \nodata  & 80.8  & \nodata & 79.4 & \nodata \\
%5341.024 & 1.6080  & -2.0065 & \nodata & \nodata  & \nodata  & \nodata &  \nodata & \nodata \nl
5373.698 & 4.474   & $-$0.8600 & \nodata & 21.2  & 15.1 & \nodata & \nodata & 61.0 \\
5379.573 & 3.695   & $-$1.4970 & 31.0 & 17.3 & \nodata & 33.6  & 17.8 & 59.3 \\
5393.167 & 3.2410  & $-$0.8125 & 75.1  & 68.5  & 69.4  & 79.8 & 72.5 & \nodata \\
%5397.128 & 0.9150  & -1.9875 & \nodata & \nodata & \nodata & \nodata & \nodata & \nodata \\
%5405.775 & 0.9900  & -1.8520 & \nodata & \nodata & \nodata & \nodata & \nodata & \nodata \\
%5429.696 & 0.9580  & -1.8800 & \nodata & \nodata & \nodata  &  \nodata& \nodata & \nodata \nl
%5434.524 & 1.0110  & -2.1220 & \nodata & \nodata & \nodata  &  \nodata& \nodata & \nodata \nl
%5455.609 & 1.0110  & -2.0945 & \nodata & \nodata & \nodata &  \nodata&  \nodata & \nodata \nl
5497.516 & 1.0110  & $-$2.8371 & 70.9  & 81.1  & 62.0  & 84.0  & 79.4 & \nodata \\
5501.465 & 0.9580  & $-$2.9978 & 77.9  & 69.6  & 62.7  & 75.4  & 72.1 & \nodata \\
5506.779 & 0.9900  & $-$2.7929 & 77.6 & 79.8  & 55.8  & 82.9  & 74.9 & \nodata \\
5560.207 & 4.435   & $-$1.1900 & \nodata & 13.9  & \nodata & \nodata  &  \nodata & \nodata \\
5569.618 & 3.4170  & $-$0.5130 & 71.3  & \nodata & \nodata & \nodata &  \nodata & 50.8 \\
5572.841 & 3.3970  & $-$0.2925 & \nodata & \nodata & 82.2  & \nodata &  \nodata & \nodata \\
5576.090 & 3.4300  & $-$1.0000 & 67.0  & 69.5  & 37.0  & 57.5  & 54.6 & \nodata \\
%5586.756 & 3.3680  & -0.1440 & \nodata & \nodata & \nodata & \nodata &  \nodata & \nodata \nl
5658.816 & 3.3970  & $-$0.8360 & 76.3  & 71.1  & \nodata & \nodata & 74.9 & \nodata \\
5717.835 & 4.285   & $-$1.1300 & 17.8 & 17.8  & 14.7  & 17.8  & \nodata & 62.6 \\
5775.080 & 4.220   & $-$1.2970 & 16.4  & 16.9  & \nodata & 16.9 & \nodata & 55.5 \\
6012.204 & 2.223   & $-$4.0381 & \nodata & \nodata  & \nodata  & \nodata & \nodata & 24.2 \\ 
6016.604 & 3.547   & $-$1.8200 & 22.4  & 14.5  & \nodata  & 15.1  & 21.4 & \nodata \\
6027.048 & 4.076   & $-$1.1495 & 27.6  & 23.9  & 15.9  & 27.8  & 26.0 & 63.4 \\
6055.992 & 4.734   & $-$0.4600 & 23.2  & 33.7  & 27.2  & 34.3  & 23.4 & 72.8 \\
6065.481 & 2.6090  & $-$1.4700 & 70.6  & 69.6  & 74.6  & 63.6  & 70.0 & \nodata \\
6078.999 & 4.652   & $-$1.1200 & \nodata & 9.6 & \nodata  & 9.6  & \nodata & 43.3 \\ 
6127.904 & 4.143   & $-$1.3990 & 17.1  & 13.1  & \nodata  & 17.6  &  6.5 & 49.7 \\
6136.615 & 2.4530  & $-$1.4050 & 78.7  & 78.4  & 71.0  & 85.8  & 82.0 & \nodata \\
6137.691 & 2.5880  & $-$1.3745 & 82.0  & 73.3  & 81.4  & 80.3  & 75.6 & \nodata \\
6151.618 & 2.176   & $-$3.2990 & 19.7  & 11.4  & 11.2  & 18.6  & \nodata & 49.0 \\
6157.725 & 4.076   & $-$1.2600 & 28.2  & 25.5  & 20.4  & 27.0  &  \nodata & 62.6 \\
6173.341 & 2.223   & $-$2.8800 & 31.6  & 19.8  & 31.4  & 40.4  & 27.6 & 68.3 \\
6180.203 & 2.728   & $-$2.6225 &  \nodata & \nodata & \nodata &  \nodata &  \nodata & 55.6 \\
6219.280 & 2.1980  & $-$2.4330 & 60.9  & 42.9  & 43.7  & 49.2  & 53.0 & 89.0 \\
6230.723 & 2.5590  & $-$1.2785 & 88.7  & 84.0  & 80.4  & 93.6  & 84.4 & \nodata \\
6240.645 & 2.223   & $-$3.2030 & \nodata & \nodata & \nodata & \nodata & \nodata & 48.0 \\ 
6246.318 & 3.6030  & $-$0.8770 & 64.0  & 60.6  & 46.4  & 58.1  & 58.7 & \nodata \\
6252.555 & 2.4040  & $-$1.7270 & 74.8  & 70.9  & 71.5  & 73.9  & 72.6 & \nodata \\
6254.257 & 2.2790  & $-$2.4430 & \nodata & 56.3  & \nodata & \nodata & \nodata & \nodata \\
6322.684 & 2.588   & $-$2.4475 & \nodata & \nodata & \nodata & \nodata & \nodata & 75.3 \\
6335.330 & 2.1980  & $-$2.2035 & 63.5  & 50.5  & 42.2  & 62.1  & 48.1 & 96.8 \\
6344.148 & 2.433   & $-$2.9000 & 17.7  & 22.4  & 18.9  & 22.4  & 18.6 & 66.4 \\
6355.029 & 2.8450  & $-$2.3575 & 28.3  & 20.2 & 25.6  & \nodata  & 16.2 & 78.1 \\
6380.742 & 4.187   & $-$1.3875 & 18.1  & 13.7  &  \nodata & 15.4  & \nodata & 51.3 \\
6393.601 & 2.4330  & $-$1.5760 & 71.8  & 68.0  & 63.4  & 72.8  & 76.0 & \nodata \\
6481.869 & 2.279   & $-$2.9840 & 24.2  & 19.4  & 22.2  & 26.9  & 18.6 & 64.1 \\
6494.980 & 2.4040  & $-$1.2560 & 96.1  & \nodata  & 90.7  & 97.3  & 88.5 & \nodata \\
6498.940 & 0.958   & $-$4.6946 & \nodata & \nodata & \nodata & \nodata & \nodata & 47.1 \\
6592.913 & 2.7280  & $-$1.5365 & 70.9  & 73.6  & 65.4  & 66.7  & 59.4 & \nodata \\
6593.868 & 2.433   & $-$2.3940 & 52.4  & 42.3  & 47.1  & 51.3  & 46.2 & 84.7 \\
6609.109 & 2.559   & $-$2.6765 & 26.2  & 17.7  & 17.7  & 17.7 & 20.7 & 67.1 \\ 
6750.152 & 2.4240  & $-$2.6080 & 28.9  & 25.6  & \nodata & 29.5  & 30.9 & 73.7 \\

%\cline{1-9}
%\cutinhead{\ion{Fe}{2} \phm{blahhhhhhhhhhhhhhhhhhhhhhhhhhhhhhhhhhhhhhhhhhhhhhhhhhhhhhhhh}} 
%\cline{1-9} \\ 
\hline
\multicolumn{1}{l}{\ion{Fe}{2}} &
\multicolumn{8}{c}{} \\
\hline
4508.289 & 2.8580  & $-$2.3210 & 63.0  & 60.3  & 41.4  & 65.4  & 55.2 & \nodata\\
4515.339 & 2.8440  & $-$2.4800 & 54.3  & 69.5  & \nodata  & 74.5  & 55.2 & \nodata\\
4522.634 & 2.8440  & $-$2.1100 & 83.3 & \nodata  & \nodata  & 85.1   & 71.4 & \nodata\\
4576.339 & 2.8410  & $-$2.9550 & 51.7  & \nodata  & 43.9  & \nodata  & 42.2 & 61.4 \\
4583.837 & 2.8070  & $-$1.9200 & 85.6  & 76.3  & 80.2  & 75.7  &  83.3 & \nodata\\
4629.339 & 2.8070  & $-$2.3700 & 64.0 & 52.8  & 73.1  & 75.9  & 58.0 & \nodata\\
4923.927 & 2.8910  & $-$1.3200 & \nodata & \nodata & 84.3  & \nodata & \nodata & \nodata\\
%5018.440 & 2.8910  &$ -$1.2250 & \nodata &  \nodata& \nodata & \nodata & \nodata & \nodata\\
5197.576 & 3.2310  & $-$2.1665 & 52.7  & 54.1  & 48.8  & 57.4  & 49.0 & 78.7 \\
5234.630 & 3.2210  & $-$2.2100 & 62.6  & 51.7  & 61.3  & 64.1  & 51.7 & \nodata\\
5276.002 & 3.2000  & $-$2.0350 & 68.0  & \nodata & 82.1 & \nodata & 74.8 & \nodata\\
5316.615 & 3.1520  & $-$2.0200 & \nodata  & 86.5  & 87.9 & \nodata  &  \nodata & \nodata\\
5325.560 & 3.221   & $-$2.0750 & \nodata  & \nodata  & \nodata  & \nodata  & \nodata & 41.7 \\
%5534.847 & 3.245   & -2.9110 & \nodata & \nodata & \nodata & \nodata & \nodata & \nodata\\ 
6149.249 & 3.890   & $-$2.9300 & 23.7  & 25.4  & \nodata  & \nodata  & \nodata & 36.6 \\
6247.562 & 3.890   & $-$2.7240 & \nodata & 22.3  & \nodata  & \nodata  & \nodata & 53.5 \\
6456.391 & 3.9000  & $-$2.3290 & 51.6  & \nodata  & 46.0  & 37.8  & 50.8 & \nodata \\

%\cline{1-9}
%\cutinhead{\ion{Ni}{1} \phm{blahhhhhhhhhhhhhhhhhhhhhhhhhhhhhhhhhhhhhhhhhhhhhhhhhhhhhhhhh}}
%\cline{1-9}\\
\hline
\multicolumn{1}{l}{\ion{Ni}{1}} &
\multicolumn{8}{c}{} \\
\hline
4714.408 & 3.3801  &  0.2300 & 67.3 & \nodata & 63.9  & \nodata & 63.6 & \nodata \\
4715.757 & 3.5435  & $-$0.3400 & 34.4 & \nodata & \nodata & 49.3  & 32.7 & 79.3 \\
4756.510 & 3.4802  & $-$0.3400 & 43.8 & 34.9 & 40.8  & 49.8  & 35.1 & 80.7 \\
4786.531 & 3.4198  & $-$0.1700 & 51.8 & 41.7 & 57.7  & 44.9  & 43.9 & 95.4 \\
4829.016 & 3.5424  & $-$0.3300 & 30.5 & 21.9 & 19.1  & 30.5  & 29.2 & 78.5 \\
4831.169 & 3.6063  & $-$0.4200 & 35.8 & 33.7 & 36.3  & 34.3  & 40.3 & 72.9 \\
4904.407 & 3.5424  & $-$0.1700 & 44.0 & 54.9 & 44.8  & 35.5  & 42.4 & 85.5 \\
4937.341 & 3.6063  & $-$0.3900 & 30.0 & 36.5 & 29.1  & 40.8  & 22.4 & 85.3 \\
5035.357 & 3.6356  &  0.2900 & 53.9 & 56.8 & 40.1  & 65.7  & 51.8 & \nodata \\
5081.107 & 3.8476  &  0.3000 & 60.9 & 61.3 & 53.4  & 52.6  & 46.0 & 95.9 \\
5084.089 & 3.6787  &  0.0300 & 54.8 & 36.4 & 47.9  & 46.5  & 47.7 & 89.5 \\
5115.389 & 3.8342  & $-$0.1100 & 41.6 & 32.1 & 37.5  & 40.6  & 34.9 & 74.6 \\
5146.480 & 3.7060  &  0.1200 & \nodata & 44.6 & 37.2  & 51.8  & \nodata & 87.0 \\
5155.762 & 3.8985  & $-$0.0900 & 44.8 & 33.7 & 27.9  & 42.0  & 21.5 & 75.1 \\
5476.900 & 1.8263  & $-$0.8900 & \nodata & \nodata & 89.9 & \nodata & 96.7 & \nodata \\
5587.853 & 1.9355  & $-$2.1400 & 30.2 & 98.6 &  \nodata& 25.0  & \nodata & 59.3 \\
%5592.259 & 1.9509  & -2.5700 & \nodata & \nodata & \nodata  & \nodata  & \nodata & \nodata \nl
%5711.883 & 1.9355  & -2.2700 & \nodata & \nodata & \nodata & \nodata & \nodata & \nodata \nl
6108.107 & 1.6765  & $-$2.4500 & \nodata & \nodata & 11.9 & \nodata & \nodata & 64.7 \\
%6256.351 & 1.6765  & -2.4800 & \nodata & \nodata & \nodata  & \nodata & \nodata & \nodata \\
6482.796 & 1.9355  & $-$2.6300 & 11.1 & 9.2  & \nodata & 10.8 & 10.2 & 41.5 \\
6586.308 & 1.9509  & $-$2.8100 & 13.9 & 11.6 & \nodata & \nodata  & \nodata & 43.3 \\ 
6643.629 & 1.6765  & $-$2.3000 & 50.7 & 40.6 & \nodata & 51.6 & 38.6 & \nodata \\
6767.768 & 1.8263  & $-$2.1700 & 47.5 & 37.4 & \nodata & 41.2 & 33.8 & \nodata \\

%\cline{1-9}
%\cutinhead{\ion{Y}{2} \phm{blahhhhhhhhhhhhhhhhhhhhhhhhhhhhhhhhhhhhhhhhhhhhhhhhhhhhhhhhh}}
%\cline{1-9}\\
\hline
\multicolumn{1}{l}{\ion{Y}{2}} &
\multicolumn{8}{c}{} \\
\hline
4883.684 & 1.0821  & $-$0.0100 & 35.2 & 40.7 & 34.1 & 35.3 & 34.9 & 55.5 \\
4900.120 & 1.0313  & $-$0.1300 & 44.6 & 50.2 & 41.2 & 41.0 & 41.6 & 58.0 \\
5087.416 & 1.0810  & $-$0.3100 & 34.1 & 33.4 & 28.3 & 22.8 & 33.9 & 43.8 \\

%\cline{1-9}
%\cutinhead{\ion{Ba}{2} \phm{blahhhhhhhhhhhhhhhhhhhhhhhhhhhhhhhhhhhhhhhhhhhhhhhhhhhhhhhhh}}
%\cline{1-9}\\
\hline
\multicolumn{1}{l}{\ion{Ba}{2}} &
\multicolumn{8}{c}{} \\
\hline
%4554.000 &  0.000  &  ?      & \nodata & \nodata & \nodata  & \nodata  & \nodata \\   
%4934.100 &  0.000  &  ?      & \nodata & \nodata & \nodata & \nodata & \nodata \\
5853.700 &  0.000  & $-$1.0000  & 48.3  & 46.7  & 53.6 & 52.7  & 41.5  & 63.3 \\
6141.700 &  0.000  & $-$0.0760  & 96.5  & 85.6  & 85.2  & 93.4  & 83.7 & 116.0 \\

\enddata
\end{deluxetable}
%\end{document}

\clearpage
%\input{tab3.tex}
%\documentclass[12pt,preprint]{aastex}
%\documentclass{aastex}
%\begin{document}

\singlespace
\begin{center}
\begin{deluxetable}{lcccc} 
\tablewidth{0pc}
\tablenum{3}
\tablecolumns{5} 
\tablecaption{Model Parameters} 
\tablehead{ 
\colhead{Star}  &  \colhead{Temperature}  & \colhead{$\log{\rm g}$}  & \colhead{$\xi$} & \colhead{Input [Fe/H]}\\
\colhead{}	&  \colhead{(K)} & \colhead{(dex)}  & \colhead{(km~s$^{-1}$)}  
  &  \colhead{(dex)} }

\startdata 
239............  & $5845\pm45$ & 4.01 & 1.00 & $-$0.70 \\
259............  & $5800\pm45$ & 3.98 & 1.00 & $-$0.70 \\
260............  & $5930\pm45$ & 4.08 & 1.10 & $-$0.70 \\
264............  & $5845\pm45$ & 4.01 & 1.00 & $-$0.70 \\
273............  & $5845\pm45$ & 4.01 & 1.00 & $-$0.70 \\
\enddata 
\end{deluxetable} 
\end{center}

%\end{document}

%% 
%% End of file `obs_table.tex'. 

\clearpage
%\input{tab4.tex}
%%Table 4
%\documentclass[12pt,preprint]{aastex}
%\documentstyle[aaspp4]{article}
%\begin{document}
%\singlespace
%\small
%\begin{center}
\begin{deluxetable}{llrrrrrrrc}
\tablewidth{0pc}
\tablenum{4}
\tablecolumns{9}
\tablecaption{Iron Abundances in  M 71}
\tablehead{
\colhead{Species} & \colhead{Quantity} & \colhead{239} & \colhead{259} & \colhead{260} & \colhead{264} & \colhead{273} & \colhead{Mean} & \colhead{Sun} & {$<$}[Fe/H]{$>$} 
}
\startdata
\ion{Fe}{1} & Mean  &  6.73     & 6.64     & 6.60     & 6.71     & 6.64     & 6.66     & 7.47    & \nodata \\
& [Fe I/H] & $-$0.74   & $-$0.83  & $-$0.87  & $-$0.76 & $-$0.83 & $-$0.81 & \nodata & $-0.81$ \\  
          & ${\sigma}_{\mu}$     &  0.016    & 0.018    & 0.018    & 0.026    & 0.016    & 0.027    & 0.019   & 0.031   \\  
           & \# lines                 &  76       & 76       & 77       & 71       & 76       & \nodata  & 32      & \nodata \\ 
           & ${\Delta}$Mean$_{\odot}$ &  $+0.044$ & $-0.047$ & $+0.060$ & $-0.027$ & $-0.008$ & $+0.005$ & \nodata & \nodata \\ 
\ion{Fe}{2} & Mean                    &  6.68     & 6.58     & 6.57     & 6.71     & 6.55     & 6.62     & 7.37    & \nodata \\  
& [Fe II/H]  & $-$0.69  & $-$0.79  & $-$0.80  & $-$0.66  & $-$0.82  & $-$0.75
 & \nodata & $-0.75$ \\
 & ${\sigma}_{\mu}$         &  0.053    & 0.096    & 0.118    & 0.078    & 0.062    & 0.037    & 0.053   & 0.065   \\ 
           & \# lines                 &  11       & 9        & 10       & 8        & 10       & \nodata  & 6       & \nodata \\  
           & ${\Delta}$Mean$_{\odot}$ &  $+0.028$ & $+0.001$ & $-0.081$ & $-0.175$ & $-0.075$ & $-0.060$ & \nodata & \nodata \\  
Fe I \& II & Wted Mean & 6.72 & 6.63 & 6.59 & 6.71 & 6.63 & 6.66 & 7.45 
& \nodata \\
& [Fe/H]  & $-$0.73  & $-$0.82  & $-$0.86  & $-$0.75  & $-$0.83  & $-$0.80 
& \nodata   & $-$0.80
\enddata
\end{deluxetable}
%\end{document}

\clearpage
%\input{tab5.tex}
%%Table 4
%\documentclass[12pt,preprint]{aastex}
%\documentstyle[aaspp4]{article}
%\begin{document}
%\singlespace
\tabletypesize{\scriptsize}
%\small
%\begin{center}
\begin{deluxetable}{llrrrrrrrc}
\tablewidth{0pc}
\tablenum{5}
\tablecolumns{9}
\tablecaption{M 71 Abundances for Na and the $\alpha$ Elements}
\tablehead{
\colhead{Species} & \colhead{Quantity} & \colhead{239} & \colhead{259} & \colhead{260} & \colhead{264} & \colhead{273} & \colhead{Mean} & \colhead{Sun} & {$<$}[X/H]{$>$} 
}
\startdata
\ion{Na}{1} & Mean & 5.53 & 5.69 & 5.75 & 5.68 & 5.61 & 5.66 & 6.32 & $-$0.66\\
& ${\sigma}_{\mu}$ & 0.08 & 0.14 & 0.10 & 0.17 & 0.06 & 0.05 & 0.02 & 0.05   \\
& \# lines         & 4    & 3    & 4    & 3    & 4 & \nodata  & 3  & \nodata \\
& ${\Delta}$Mean$_{\odot}$ & $+0.067$  & $+0.102$ & $+0.076$ & $+0.100$  & $+0.049$ & $+0.078$ & \nodata & \nodata \\ 
& [Na/H]  & $-$0.79 & $-$0.63 & $-$0.57 & $-$0.64 & $-$0.71 &$-$0.66 &\nodata
&$-$0.66 \\
\ion{Mg}{1} & Mean & 6.90 & 6.80 & 6.79 & 6.78 & 6.90 & 6.83 & 7.54 & $-0.71$\\
& ${\sigma}_{\mu}$ & 0.06 & 0.08 & 0.42 & 0.05 & 0.18 & 0.03 & 0.14 & 0.14 \\  
& \# lines         & 3    & 3    & 2    & 2    & 3 & \nodata & 3  & \nodata \\
& ${\Delta}$Mean$_{\odot}$ &  $-0.082$  & $-0.077$ & $+0.390$ & $+0.125$ & $-0.013$ & $+0.069$ & \nodata & \nodata \\
& [Mg/H]  & $-$0.64 & $-$0.74 & $-$0.75 & $-$0.76 & $-$0.64 &$-$0.71 &\nodata
&$-$0.71 \\
\ion{Si}{1} & Mean & 7.08 & 7.06 & 6.93 & 7.13 & 7.00 & 7.04 & 7.61 & $-0.57$\\
& ${\sigma}_{\mu}$ & 0.06 & 0.12 & 0.13 & 0.13 & 0.07 & 0.04 & 0.03 & 0.05 \\
& \# lines         & 3    & 3    & 3    & 3    & 3 & \nodata  & 3 & \nodata \\
\ion{Si}{2} & Mean & 6.95 & 6.82 & 6.76 & 6.99 & 7.01 & 6.91 & 7.55 & $-0.64$\\
& ${\sigma}_{\mu}$ & \nodata & \nodata & \nodata & \nodata & \nodata & 0.06  
& \nodata & 0.07  \\  
& \# lines         & 1    & 1    & 1   & 1     & 1 & \nodata  & 1 & \nodata \\ 
Si I \& II & Wted Mean & 7.05 & 7.00 & 6.89 & 7.10 & 7.00 & 7.01 & 7.59 & \nodata \\
& [Si/H]  & $-$0.54 & $-$0.59 & $-$0.70 & $-$0.49 & $-$0.59 &$-$0.58 &\nodata
&$-$0.58\\
\ion{Ca}{1} & Mean & 5.78 & 5.62 & 5.68 & 5.77 & 5.63 & 5.70 & 6.18 & $-$0.48\\
& ${\sigma}_{\mu}$ & 0.03 & 0.03 & 0.04 & 0.03 & 0.04& 0.04 & 0.03 & 0.04 \\  
& \# lines         & 20   & 19   & 18   & 19   & 18 & \nodata & 21 & \nodata\\
& ${\Delta}$Mean$_{\odot}$ & $-0.027$  & $+0.018$ & $-0.028$ &
$-0.010$ & $+0.004$ & $-0.009$ & \nodata & \nodata \\ 
& [Ca/H]  & $-$0.40 & $-$0.56 & $-$0.50 & $-$0.41 & $-$0.55 &$-$0.48 &\nodata
&$-$0.48 \\
\ion{Ti}{1} & Mean & 4.47 & 4.28 & 4.41 & 4.58 & 4.28 & 4.40 & 4.83 & $-0.43$\\
& ${\sigma}_{\mu}$ & 0.03 & 0.04 & 0.05 & 0.04 & 0.04 & 0.05 & 0.03 & $0.06$\\
& \# lines         & 14   & 14   & 13   & 13   & 13  & \nodata & 8& \nodata\\ 
& ${\Delta}$Mean$_{\odot}$ & $-0.004$  & $+0.037$ & $-0.025$ &$-0.011$ &
$-0.023$ & $-0.005$ & \nodata & \nodata \\  
\ion{Ti}{2} & Mean & 4.52 & 4.44 & 4.36 & 4.68 & 4.40 & 4.48 & 5.09& $-0.61$\\
& ${\sigma}_{\mu}$ & 0.028& 0.058& 0.082&0.079& 0.041& 0.064&0.067& 0.086 \\  
& \# lines         & 7    & 9    & 9    & 12  & 9 & \nodata & 6  & \nodata \\  
& ${\Delta}$Mean$_{\odot}$ & $+0.012$  & $-0.086$ & $+0.079$ & $+0.034$  & $+0.022$ & $+0.018$ & \nodata & \nodata \\ 
Ti I \& II & Wted Mean & 4.49 & 4.34 & 4.39 & 4.62 & 4.33 & 4.43 & 4.94 & \nodata \\
& [Ti/H] & $-$0.45 & $-$0.55 & $-$0.55 & $-$0.32 & $-$0.61 & $-$0.50 & \nodata
& $-$0.50 \\
\enddata
\end{deluxetable}
%\end{document}

\clearpage
%\input{tab6.tex}
%%Table 4
%\documentclass[12pt,preprint]{aastex}
%\documentstyle[aaspp4]{article}
%\begin{document}
%\singlespace
\small
%\begin{center}
\begin{deluxetable}{llrrrrrrrc}
\tablewidth{0pc}
\tablenum{6}
\tablecolumns{9}
\tablecaption{M 71 Abundances for Fe-Peak and N-Capture Elements}
\tablehead{
\colhead{Species} & \colhead{Quantity} & \colhead{239} & \colhead{259} & \colhead{260} & \colhead{264} & \colhead{273} & \colhead{Mean} & \colhead{Sun} & {$<$}[X/H]{$>$} 
}
\startdata

\ion{Ni}{1} & Mean &  5.44     & 5.30     & 5.29     & 5.37     & 5.29     & 5.34     & 6.05    & $-0.71$ \\ 
& ${\sigma}_{\mu}$ &  0.04    & 0.04    & 0.05    & 0.04    & 0.03    & 0.03    & 0.03   & $0.04$ \\  
& \# lines &  18       & 17       & 15       & 17       & 17       & \nodata  & 16      & \nodata \\ 
& ${\Delta}$Mean$_{\odot}$ &  $-0.007$ & $-0.012$ & $+0.027$ & $-0.023$ & $-0.018$ & $-0.007$ & \nodata & \\ 
& [Ni/H] & $-$0.61  & $-$0.75 & $-$0.76 & $-$0.68 & $-$0.76 & $-$0.71 &
\nodata & $-$0.71 \\
\ion{Cr}{1} & Mean & 4.76      & 4.79     & 4.80     & 5.01     & 4.72     & 4.81     & 5.53    & $-0.72$ \\   
& ${\sigma}_{\mu}$ & 0.03     & 0.04    & 0.06    & 0.05    & 0.03    & 0.06    & 0.04   & $0.07$ \\  
& \# lines & 13        & 14       & 12       & 12       & 12       & \nodata  & 15      & \nodata \\  
& ${\Delta}$Mean$_{\odot}$ & $+0.003$  & $+0.023$ & $-0.029$ &
           $-0.026$ & $-0.009$ & $-0.008$ & \nodata & \nodata \\  

\ion{Cr}{2} & Mean & 4.79      & 4.88     & 4.76     & 4.89     & 4.90     & 4.84     & 5.73    & $-0.89$ \\  
& ${\sigma}_{\mu}$ & 0.04     & 0.04    & 0.11    & 0.11    & 0.10    & 0.03    & 0.09   & $0.09$ \\  
& \# lines  & 4         & 5        & 5        & 4        & 5        & \nodata  & 5       & \nodata \\  

Cr I \& II & Wted Mean & 4.77 & 4.81 & 4.79 & 4.98 & 4.78 & 4.83 & 5.58 &
\nodata \\
& [Cr/H] & $-$0.81  & $-$0.77 & $-$0.79 & $-$0.60 & $-$0.80 & $-$0.75 &
\nodata & $-$0.75 \\

\ion{Y}{2}  & Mean & 1.53      & 1.60     & 1.48     & 1.40     & 1.51     & 1.50     & 2.15    & $-0.65$ \\
& ${\sigma}_{\mu}$  & 0.11     & 0.11    & 0.09    & 0.10    & 0.10    & 0.04    & 0.05   & 0.06   \\
& \# lines    & 3         & 3        & 3        & 3        & 3        & \nodata  & \nodata & \nodata \\
& [Y/H] & $-$0.62  & $-$0.55 & $-$0.67 & $-$0.75 & $-$0.64 & $-$0.65 &
\nodata & $-$0.65 \\

\ion{Ba}{2} & Mean & 1.63      & 1.49     & 1.61     & 1.67     & 1.47     & 1.57     & 2.14    & $-0.57$ \\
& ${\sigma}_{\mu}$ & \nodata   & \nodata  & \nodata  & \nodata  & \nodata  & 0.045    & \nodata & 0.05    \\
& \# lines & 2         & 2        & 2        & 2        & 2        & \nodata  & 2       & \nodata \\
& [Ba/H] & $-$0.51  & $-$0.65 & $-$0.53 & $-$0.47 & $-$0.67 & $-$0.57 &
\nodata & $-$0.57 \\

\enddata
\end{deluxetable}
%\end{document}

\clearpage
%\input{tab7.tex}
%%\documentclass{aastex}
%\documentclass[12pt,preprint]{aastex}
%\documentstyle[aaspp4]{article}
%\begin{document}

%\singlespace
\begin{center}
\begin{deluxetable}{lcccl} 
\tablewidth{0pc}
\tablenum{7}
\tablecolumns{5} 
\tablecaption{Cluster Abundance Summary} 
\tablehead{ 
\colhead{Element Ratio}  &  \colhead{Abundance}  & \colhead{$\sigma$}  &
\colhead{\# lines/star} & \colhead{Species}
}
\startdata 
${\rm [Fe/H]}$   & $-$0.80 & $\pm$0.02 & 79-86 & \ion{Fe}{1} and \ion{Fe}{2} \\
${\rm [Na/Fe]}$  & +0.14   & $\pm$0.04 & 3-4   & \ion{Na}{1} \\
${\rm [Mg/Fe]}$  & +0.09   & $\pm$0.03 & 2-3   & \ion{Mg}{1}  \\
${\rm [Si/Fe]}$  & +0.22   & $\pm$0.04 & 4     & \ion{Si}{1} and \ion{Si}{2} \\
${\rm [Ca/Fe]}$  & +0.32   & $\pm$0.04 & 18-20 & \ion{Ca}{1}  \\
${\rm [Ti/Fe]}$  & +0.30   & $\pm$0.06 & 21-25 & \ion{Ti}{1} and \ion{Ti}{2} \\
${[< \alpha >/{\rm Fe}]}$ & +0.29 & $\pm$0.05 & 45-52 & Mg, Si, Ca, Ti \\
${\rm [Cr/Fe]}$  & +0.05   & $\pm$0.04 & 16-19 & \ion{Cr}{1} and \ion{Cr}{2} \\
${\rm [Ni/Fe]}$  & +0.09   & $\pm$0.03 & 15-18 & \ion{Ni}{1}  \\
${\rm [Y/Fe]}$   & +0.15   & $\pm$0.04 & 3     & \ion{Y}{2}  \\
${\rm [Ba/Fe]}$  & +0.23   & $\pm$0.04 & 2     & \ion{Ba}{2}  \\
\enddata 
\end{deluxetable} 
\end{center}

%\end{document}

%% 
%% End of file `tab7.tex'. 

\clearpage
%\input{tab8.tex}
%\documentclass[12pt,preprint]{aastex}
%\begin{document}
\begin{deluxetable}{lrrcccc}
\tablewidth{0pc}
\tablenum{8}
\tablecolumns{7}
\tablecaption{Mean Cluster Abundances and Parameter Sensitivities}
\tablehead{ 
\colhead{Ratio} & \colhead{Mean} & \colhead{{$\sigma$}$_{\mu}$(int)} & \colhead{{$\sigma$}$_{\mu}$(int$+{\odot}$)}  & \colhead{${\Delta}T_{\rm eff}$} & \colhead{{$\Delta$}log $g$} & \colhead{{$\Delta$}$\xi$} \\ 
   &  & \colhead{dex} & \colhead{dex} & \colhead{${\pm}100$ K} & \colhead{${\pm}0.3$ dex} & \colhead{${\pm}0.2$ km/s} 
}
\startdata 
${\rm [Fe/H]}$  & $-0.80$ & 0.025 & 0.035 & ${\pm}0.074$ & ${\pm}0.005$ & 
${\mp}0.052$ \\ 
${\rm [Na/Fe]}$ & $+0.14$ & 0.064 & 0.085 & ${\mp}0.022$ & ${\mp}0.020$ &
${\pm}0.050$ \\ 
${\rm [Mg/Fe]}$ & $+0.09$ & 0.028 & 0.12? & ${\pm}0.059$ & ${\mp}0.005$ & 
${\pm}0.012$ \\  
${\rm [Si/Fe]}$ & $+0.22$ & 0.015 & 0.046 & ${\mp}0.053$ & ${\pm}0.006$ & 
${\pm}0.038$ \\  
${\rm [Ca/Fe]}$ & $+0.32$ & 0.023 & 0.043 & ${\mp}0.003$ & ${\mp}0.042$ & 
${\pm}0.025$ \\  
${\rm [Ti/Fe]}$ & $+0.30$ & 0.032 & 0.056 & ${\pm}0.008$ & ${\pm}0.021$ &
 ${\mp}0.001$ \\  
${\rm [Cr/Fe]}$ & $+0.05$ & 0.044 & 0.068 & ${\mp}0.004$ & ${\pm}0.016$ & 
${\pm}0.015$ \\  
${\rm [Ni/Fe]}$ & $+0.09$ & 0.013 & 0.040 & ${\pm}0.005$ & ${\mp}0.022$ & 
${\pm}0.024$ \\  
${\rm [Y/Fe]}$  & $+0.15$ & 0.051 & 0.076 & ${\mp}0.047$ & ${\pm}0.101$ & 
${\pm}0.006$ \\  
${\rm [Ba/Fe]}$ & $+0.23$ & 0.037 & 0.11? & ${\mp}0.021$ & ${\pm}0.035$ & 
${\mp}0.028$ \\ 
\enddata

\end{deluxetable}
%\end{document}

\clearpage
%\input{tab9.tex}
%\documentclass[12pt,preprint]{aastex}
%\begin{document}

%\singlespace
\begin{center}
\begin{deluxetable}{lcccccccccc} 
\tabletypesize{\footnotesize}
\tablewidth{0pc}
\tablenum{9}
\tablecolumns{11} 
\tablecaption{Data on Comparison Stars}
\tablehead{ 
\colhead{Star}  & \colhead{[Fe/H]}  & \colhead{[Na/Fe]}  & \colhead{[Mg/Fe]}  & \colhead{[Si/Fe]}  &
\colhead{[Ca/Fe]}  & \colhead{[Ti/Fe]} & \colhead{[Cr/Fe]}  & \colhead{[Ni/Fe]}  & \colhead{[Y/Fe]}  & \colhead{[Ba/Fe]}
}
%STAR     &[Fe/H]  &[Na/Fe] &[Mg/Fe] &[Si/Fe] &[Ca/Fe] &[Ti/Fe] &[Cr/Fe] &[Ni/Fe] &[Y/Fe] &[Ba/Fe]  \\   

\startdata 
\multicolumn{3}{l}{Stars from SB02} &
\multicolumn{8}{c}{} \\
\hline
G33-31	&$-$1.12 &$-$0.16 &+0.29  &$+$0.17  &+0.27 &+0.27  &$+$0.03  &$-$0.04 &$+$0.06 &+0.12 \\ 
G5-19	&$-$1.16 &$-$0.21 &+0.20  &$+$0.12  &+0.19 &+0.22  &$+$0.00  &$-$0.10 &$-$0.05 &+0.15 \\
G82-5	&$-$0.71 &$-$0.48 &+0.07  &$-$0.06  &+0.01 &+0.08  &$+$0.00  &$-$0.13 &$-$0.03 &+0.17 \\ 
G9-36	&$-$1.12 &$-$0.06 &+0.32  &$+$0.18  &+0.23 &+0.22  &$+$0.04  &$-$0.07 &$+$0.04 &+0.17 \\ 
G114-42	&$-$1.11 &$-$0.20 &+0.37  &$+$0.20  &+0.21 &+0.19  &$+$0.01  &$-$0.06 &$-$0.07 &+0.14 \\ 
G116-53	&$-$1.03 &$-$0.27 &+0.16  &$+$0.12  &+0.16 &+0.19  &$+$0.00  &$-$0.09 &$-$0.10 &+0.21 \\ 
G121-12	&$-$0.80 &$-$0.29 &+0.06  &$-$0.02  &+0.10 &+0.13  &$-$0.05  &$-$0.22 &$-$0.10 &+0.15 \\ 
G168-42	&$-$0.84 &$-$0.14 &+0.27  &$+$0.08  &+0.11 &+0.26  &$+$0.11  &$-$0.11 &$+$0.02 &+0.13 \\ 
MEAN    &$-$0.99 &$-$0.23 &+0.22  &$+$0.10  &+0.16 &+0.20  &$+$0.02  &$-$0.10 &$-$0.03 &+0.16 \\ 
\hline
\multicolumn{3}{l}{Stars from EAGLNT} &
\multicolumn{8}{c}{} \\
\hline
HR 2883   &$-$0.75 &+0.03 &+0.29 &+0.15 &+0.17 &+0.21 &\nodata &+0.03 &$+$0.01 &$-$0.04 \\   
HR 8181   &$-$0.67 &+0.14 &+0.14 &+0.07 &+0.05 &+0.09 &\nodata &+0.00 &$-$0.03 &$-$0.03 \\   
HR 3018   &$-$0.78 &+0.18 &+0.38 &+0.21 &+0.26 &+0.14 &\nodata &$-$0.03 &$+$0.01 &$-$0.05 \\   
HR 4657   &$-$0.70 &+0.12 &+0.34 &+0.23 &+0.19 &+0.32 &\nodata &+0.06 &$+$0.01 &$-$0.13 \\   
HD 22879  &$-$0.84 &+0.03 &+0.44 &+0.19 &+0.19 &+0.19 &\nodata &+0.01 &$+$0.04 &$-$0.09 \\   
HD 25704  &$-$0.85 &+0.10 &+0.35 &+0.15 &+0.24 &+0.27 &\nodata &$-$0.02 &$+$0.00 &$-$0.03 \\   
HD 51929  &$-$0.64 &+0.07 &+0.32 &+0.16 &+0.14 &+0.27 &\nodata &+0.03 &$-$0.05 &$-$0.11 \\   
HD 62301  &$-$0.69 &+0.03 &+0.30 &+0.18 &+0.07 &+0.23 &\nodata &+0.08 &$-$0.08 &$-$0.09 \\   
HD 78747  &$-$0.64 &+0.12 &+0.34 &+0.19 &+0.20 &+0.30 &\nodata &+0.03 &$+$0.07 &$+$0.00 \\   
HD 130551 &$-$0.62 &+0.17 &+0.18 &+0.10 &+0.10 &+0.20 &\nodata &+0.01 &$-$0.01 &$+$0.02 \\   
HD 134169 &$-$0.83 &+0.15 &+0.31 &+0.19 &+0.15 &+0.29 &\nodata &+0.04 &$-$0.07 &$-$0.16 \\   
HD 148211 &$-$0.65 &+0.05 &+0.30 &+0.15 &+0.15 &+0.21 &\nodata &+0.02 &$-$0.10 &$-$0.13 \\   
HD 148816 &$-$0.74 &+0.13 &+0.32 &+0.18 &+0.15 &+0.27 &\nodata &+0.02 &$-$0.03 &$-$0.15 \\   
HD 159307 &$-$0.71 &+0.17 &+0.24 &+0.14 &+0.12 &+0.09 &\nodata &+0.03 &$+$0.00 &$-$0.03 \\   
HD 210752 &$-$0.64 &+0.03 &+0.16 &+0.04 &+0.07 &+0.06 &\nodata &$-$0.03 &$-$0.19 &$-$0.14 \\   
HD 215257 &$-$0.65 &+0.09 &+0.10 &+0.02 &+0.10 &+0.17 &\nodata &$-$0.02 &$-$0.06 &$-$0.05 \\   
HD 218504 &$-$0.62 &+0.04 &+0.29 &+0.14 &+0.13 &+0.17 &\nodata &+0.02 &$-$0.08 &$-$0.10 \\   
MEAN      &$-$0.71 &+0.10 &+0.28 &+0.15 &+0.12 &+0.20 &\nodata &+0.02 &$-$0.03
&$-$0.08 \\
\hline

\enddata 
\end{deluxetable} 
\end{center}

%\end{document}

\end{document}